\documentclass[aps,prb,twocolumn,superscriptaddress,a4paper,english,longbibliography]{revtex4-1}
\usepackage{silence}
\usepackage{times}
\usepackage{color}
\usepackage[colorlinks=true,urlcolor=blue,citecolor=blue]{hyperref}
\usepackage{url}
\usepackage{soul}
\usepackage{graphicx}
\usepackage{amsmath}
\usepackage{subfigure}
\usepackage{physics}
\usepackage{setspace}
\usepackage{xcolor}
\usepackage{amssymb}
\usepackage{latexsym}
\usepackage{hyperref}
\DeclareGraphicsExtensions{.png,.eps}
\usepackage{float}
\usepackage{ulem}
\usepackage{appendix}
\usepackage{etoolbox}
\usepackage{fancyhdr}

\usepackage{xcolor}
\usepackage[urlcolor=blue]{hyperref}      
\hypersetup{
    colorlinks = true,                    
    citecolor = {blue},
    linkcolor = {purple},
           }

\begin{document}

\title{Solving and Completing the Rabi-Stark Model in the Ultrastrong Coupling Regime}

\author{Gen Li}
\affiliation{School of Physics, Beihang University,100191,Beijing, China}

\author{Hao Zhu}
\affiliation{School of Physics, Beihang University,100191,Beijing, China}
\author{Guo-Feng Zhang}
\email[Corresponding author: ]{gf1978zhang@buaa.edu.cn}
\affiliation{School of Physics, Beihang University,100191,Beijing, China}



\begin{abstract}
	In this work, we employ a unitary transformation with a suitable parameter to convert the quantum Rabi-Stark model into a Jaynes-Cummings-like model. Subsequently, we derive the analytical energy spectra in the ultrastrong coupling regime. The energy spectra exhibit a phenomenon known as spectral collapse, indicating the instability of the model due to the unboundedness of its energy from below at higher coupling parameters. To stabilize the Rabi-Stark model, we introduce a nonlinear photon-photon interaction term. We then compare the modified model with the original model in the classical oscillator (CO) limit. Interestingly, we observe a regular ``staircase" pattern in the mean photon number of the ground state. This pattern exhibits a fixed slope and equal step width, which we determine analytically.  Moreover, we analytically determine the phase boundary, which slightly differs from that in the original Rabi-Stark model. These findings offer insights into the investigation of those superradiant phase transitions that are unbounded from below due to the phenomenon of spectral collapse.
\end{abstract}
\maketitle


\section{INTRODUCTION} 


The quantum Rabi model~\cite {PhysRev.49.324, PhysRev.51.652}, which describes the interaction between a two-level system and a single quantized harmonic oscillator, has been extensively studied in various fields such as trapped ion systems~\cite {PhysRevX.8.021027}, cold atoms~\cite {PhysRevA.98.021801}, and cavity quantum electrodynamics~\cite {10.1093}. Its Hamiltonian is given by
\begin{equation}
\begin{split}
	\hat{H}_\text{R} = \omega a^{\dagger} a + \frac{\Delta}{2}\sigma_z + g(a^{\dagger} + a)\sigma_x ,
\end{split}
\end{equation}
where $ a $ and $ a^{\dagger} $ are the annihilation and creation operators of the oscillator with frequency $\omega$, respectively. $\Delta$ represents the resonant frequency of the two-level system, $ \sigma_i $ (with $i = x, y, z$) denotes the Pauli operators, and $g$ is the linear interaction strength. The presence of counter-rotating wave terms makes the exact solution of the quantum Rabi model challenging. To address this difficulty, Jaynes and Cummings~\cite {1443594} introduced the rotating wave approximation (RWA) and established a connection between the quantum Rabi model and the idealized Jaynes-Cummings model which, with U(1) symmetry, can be easily solved by considering a finite-dimensional subspace of the Hilbert space. However, the RWA may lead to a loss of accuracy  especially in large coupling region when applied to the quantum Rabi model. To overcome these limitations and retain the physical information of the Hamiltonian, various transformation methods have been proposed. By employing a series of transformations and operations, it is possible to convert the original Rabi model into a Jaynes-Cummings-like model under specific conditions~\cite {PhysRevA.86.015803}. This approach provides an alternative method for solving the quantum Rabi model, instead of relying solely on the rotating wave approximation.

Several models, derived from the quantum Rabi model, exhibit distinct physical properties compared to the original Rabi model~\cite {PhysRevA.87.033814, PhysRevLett.124.073602, PhysRevLett.109.229901, 1981PhLA...81..132B}. One such model is the quantum Rabi-Stark model, which incorporates a freely adjustable nonlinear term. The Rabi-Stark model has been extensively studied in theoretical research, revealing numerous novel features ~\cite {PhysRevA.102.063721, Maciejewski_2015, Eckle_2017, Xie_2017, Xie_2019}. Notably, this model exhibits a quantum phase transition when a specific critical value of the model's parameter is exceeded~\cite {PhysRevA.102.063721}. Traditionally, the study of quantum phase transitions in infinite-component systems requires the thermodynamic limit.\textbf{ However, recent research ~\cite {PhysRevA.106.023705, PhysRevLett.129.183602, PhysRevA.85.043821, PhysRevLett.115.180404, PhysRevLett.117.123602} has shown that the classical oscillator (CO) limit can replace the thermodynamic limit in finite-component systems. This is because both the CO limit and the thermodynamic limit realize the same mean-field transition when considering a quantum phase transition through mean-field theory~\cite {PhysRevA.85.043821}. In the CO limit, characterized by an infinite ratio of qubit frequency to field frequency, quantum phase transitions can be observed even in the simplest quantum Rabi model and the Jaynes-Cummings model~\cite {PhysRevLett.115.180404, PhysRevLett.117.123602}.} These findings contributed to a deeper understanding of quantum phase transitions.

Apart from the novel phase transition features of the Rabi-Stark model, its energy spectrum is equally captivating. In the ultrastrong coupling regime, characterized by a relatively large value of the interaction strength $ g $~\cite {Pedernales_2015} that has been experimentally achieved~\cite {Niemczyk2010, PhysRevLett.105.237001, Askenazi_2014, PhysRevA.96.012325, Forn-Diaz2017, PhysRevB.79.201303}, the Rabi-Stark model exhibits the phenomenon of spectral collapse~\cite {PhysRevA.95.053854, PhysRevA.92.033817} at a critical point where the entire negative energy level collapses. Upon crossing this critical point, the system's lowest energy decreases indefinitely, rendering the Hamiltonian unbounded from below and revealing incomplete physical characteristics~\cite {Cordeiro_2007}. While numerous studies have primarily focused on the energy spectrum properties within the well-defined parameter regime of this model, further investigation is required to understand the instability induced by the absence of the ground state in the Rabi-Stark model.

In this paper, we employ a similar methodology to solve the quantum Rabi model by simplifying the Rabi-Stark model into a Jaynes-Cummings-like model using an appropriate parameter. We compare our obtained results with numerical simulations, where the Hilbert space is truncated to a finite dimension, and we also investigate the sources of error. To address the incomplete characteristics of the Rabi-Stark model, we introduce a nonlinear photon-photon interaction term to eliminate the spectral collapse, thereby stabilizing the model~\cite {Cordeiro_2007, PhysRevA.102.033334}. Furthermore, we conduct additional investigations to explore the physical properties of the modified Rabi-Stark model and analyze the impact of the newly introduced term in comparison to the original model.

The structure of this paper is organized as follows: In Section II, we introduce the Rabi-Stark model along with our proposed completed Rabi-Stark model. The method employed to solve both models is presented in Section III, where the obtained results are also provided. Finally, the conclusions are drawn in Section IV.

\section{MODELS}
As one of the simplest and most fundamental theoretical models in quantum optics, the quantum Rabi model holds significant influence on the development of other models. The approach used to solve the quantum Rabi model can also be applied, to some extent, in solving these related models. In this section, we introduce the models under study, which share similarities with the quantum Rabi model.

\subsection{Rabi-Stark Model}
Grimsmo and Parkins~\cite {PhysRevA.87.033814} proposed a scheme in which the system comprises two stable hyperfine ground states of a multilevel atom, an optical cavity mode, and two additional laser fields. This scheme offers a generalization of the quantum Rabi model, referred to as the Rabi-Stark model, by introducing an extra nonlinear term to its effective Hamiltonian. The corresponding Hamiltonian for the Rabi-Stark model is given by
\begin{equation}\label{yuanshimao}
\begin{split}
	\hat{H}_\text{RS} =& \hat{H}_\text{R}+ \frac{U}{2} a^{\dagger}a\sigma_z\\
	=& \omega a^{\dagger}a +  \frac{\Delta}{2}\sigma_z+ g\left( a^{\dagger}+a\right)\sigma_x  + \frac{U}{2} a^{\dagger}a\sigma_z .
\end{split}
\end{equation}
In this model, the first three terms align with the quantum Rabi model. The final term in the Hamiltonian represents the nonlinear coupling between the atom and the field. Here, $U$ denotes the interaction strength associated with the dynamical Stark shift, which serves as the quantum counterpart of the classical Bloch-Siegert shift. Notably, in the experimental arrangement of the aforementioned quantum Rabi-Stark model, the nonlinear coupling strength $ U $ is freely adjustable, distinguishing it from the typical dynamical Stark shift.

\subsection{Completed Rabi-Stark Model}

As mentioned in the introduction and elaborated upon in the following section, the Rabi-Stark model is prone to instability due to the spectral collapse phenomenon occurring when the nonlinear coupling strength $U$ exceeds the critical value of $2\omega$. This phenomenon suggests that the system's energy becomes unbounded from below for large $U$ values, resulting in the absence of a ground state—a highly counterintuitive spectral feature.

Motivated by the concept of the completed Buck-Sukumar model~\cite {Cordeiro_2007} and the quest for potential stabilization methods, we introduce a variant of the Rabi-Stark model named as the completed Rabi-Stark model. In this model, we incorporate a nonlinear photon term $ \kappa(a^{\dagger} a )^2 $, specifically, a photon-photon interaction term. This addition enables the generation of topological photon pairs with robust transport properties~\cite {PhysRevA.102.033334}, which have attracted attention in experimental study~\cite {PhysRevA.95.053866, Mittal2018, Olekhno2020}. The modified Hamiltonian is given by the expression 
\begin{equation}\label{wanzhengmao}
\begin{split}
	&\hat{H}_\text{cRS} = \hat{H}_\text{RS} + \kappa (a^{\dagger} a)^2 =  \hat{H}_{\text{R}} + \frac{U}{2} a^{\dagger} a \sigma_z  + \kappa (a^{\dagger} a)^2\\
	& = \omega a^{\dagger} a +  \frac{\Delta}{2}\sigma_z + g\left( a^{\dagger}+a\right)\sigma_x  + \frac{U}{2} a^{\dagger}a\sigma_z + \kappa (a^{\dagger} a)^2  .
\end{split}
\end{equation}
In the presented Hamiltonian expression, the final term represents the interaction between photons. The coupling parameter $ \kappa $, governing the nonlinear photon-photon interaction, is small enough to preserve the unique physical characteristics of the original model. Introducing this nonlinear photon-photon interaction term, we demonstrate that the modified model successfully mitigates the occurrence of spectral collapse and exhibits notable deviations from the original model in terms of both energy spectrum and phase transition.

To ensure clarity and precision, we assign specific names to various coupling parameters in this article. The linear interaction strength, denoted by $g$, is referred to as the Rabi coupling. The nonlinear coupling strength, denoted by $U$, is referred to as the Stark coupling. Finally, the parameter $\kappa$ governing the nonlinear photon coupling is referred to as the photon coupling.

\section{METHOD AND RESULTS}

The organization of this section is as follows: In Section A, we use an approximation method to solve the quantum Rabi-Stark model under specific conditions, obtaining an approximate analytical expression for its energy in the ultrastrong coupling regime. Section B focuses on the Rabi-Stark model in the CO limit, where quantum phase transitions commonly occur. In Section C, we stabilize the quantum Rabi-Stark model by introducing a nonlinear photon-photon interaction term and discuss the energy spectra properties of the completed Rabi-Stark model. Section D explores the completed Rabi-Stark model in the CO limit, studying its physical features and comparing them with the original Rabi-Stark model.


\subsection{Energy Spectrum}

We first start with a rotation of the Rabi-Stark Hamiltonian [Eq.(\ref{yuanshimao})] around the $y$-axis, the Hamiltonian becomes
\begin{eqnarray}\label{xuanzhuanmao}
	\hat{H}_\text{E} = \left( \frac{\Delta}{2}+\frac{U}{2}a^{\dagger}a\right) \sigma_x+\omega a^{\dagger}a-g\left( a^{\dagger}+a\right)\sigma_z.
\end{eqnarray}
Then we perform a unitary transformation with $ \hat{A} = \exp[\lambda \sigma_z(a^{\dagger} - a)] $ and transform the Hamiltonian into the representation of $ \sigma_x $ (i.e., $ \sigma_x\ket{\pm x} = \pm\ket{\pm x} $ and $ \sigma_x = \tau_z $). When the parameter $ \lambda $ is chosen to satisfy that
\begin{equation}
\begin{split}
	\frac{\lambda\omega + g}{\lambda e^{-2\lambda^2}} + \frac{\Delta + U \lambda^2 + U n}{n + 1}  L_{n}^1(4\lambda^2) = \frac{U}{2} L_n(4\lambda^2) T_z,
\end{split}
\end{equation}
where $ T_z $ is a real parameter that takes on value of $ \pm 1 $. The detailed derivations can be found in Appendix A. The effective Hamiltonian takes the form of a  Jaynes-Cummings-like model, which is predominantly diagonal except for its final term in the Hamiltonian,
\begin{equation}\label{hamidun}
\begin{split}
	H_\text{E} = \omega a^{\dagger} a + \frac{\tilde{\Delta}}{2} \tau_z  + \frac{U}{2} \tau_z f(a^{\dagger} a) + \tilde{C} + \tilde{g} (\tau_+ a + \tau_- a^{\dagger} ), 
\end{split}
\end{equation}
where
\begin{equation}
\begin{split}
	f(a^{\dagger} a) &= e^{-2\lambda^2} L_n(4\lambda^2)a^{\dagger} a \\
	&+ 2\lambda^2 e^{-2\lambda^2} \left[ \frac{L_{n}^1(4\lambda^2) a^{\dagger}a }{n + 1} - \frac{a L_{n + 1}^1 (4\lambda^2) a^{\dagger}}{n + 2} \right] ,\\
	\tilde{\Delta} &= (\Delta + U \lambda^2) L_{n}(4 \lambda^2) e^{-2\lambda^2},\\
	\tilde{C} &= 2 g \lambda + \omega \lambda^2,\\
	\tilde{g} &= \lambda \omega + g - \frac{ \lambda \left(\Delta + U\lambda^2 + Un\right) L_{n}^1(4\lambda^2) e^{-2\lambda^2} }{n + 1} \\
	& - \frac{U \lambda}{2} L_n(4\lambda^2) e^{-2\lambda^2} T_z .
\end{split}
\end{equation}
From the Hamiltonian expression, the energy spectra can be obtained within the subspace $\{\ket{+x,n}, \ket{-x,n+1}\}$, $ n = 0,1,2\cdots  $, given by
\begin{equation}\label{erchenger}
	H_\text{E} = 
	\left( \begin{array}{c c c}
		H_{11} & H_{12}\\
		H_{21} & H_{22}
	\end{array}
	\right).
\end{equation}

In the experimental setup where the Rabi coupling $ g $ is relatively small($g < 0.5 \omega $) in the ultrastrong coupling regime, the parameter $\lambda$ is also small ~\cite {PhysRevA.86.015803}. The Laguerre polynomial and the associated Laguerre polynomial can be expanded up to the zero-order term, $L_n(4\lambda^2) \simeq 1$ and $L_n^1(4\lambda^2) \simeq n + 1$, respectively, leading to an approximate analytical solution
\begin{equation}\label{huajianhou}
	\begin{split}
	&\lambda \simeq \\ 
	&-\frac{g}{\omega+(\mathrm{\Delta} \pm \frac{U}{2}+Un)\exp{\left[-2\left(\frac{g}{\omega+\mathrm{\Delta} \pm U/2+Un}\right)^2\right]}}.
	\end{split}
\end{equation}
Furthermore, since the Hamiltonian [Eq.(\ref{hamidun})] adopts the Jaynes-Cummings form, the energy expectation value of $ \ket{-x, 0} $ can be directly obtained as
\begin{align}\label{gnd-eng}
	E_\text{0} =\omega\lambda^2+2\lambda\ g - \frac{\Delta - U \lambda^2+4U\lambda^4}{2}e^{-2\lambda^2},
\end{align}
the state $ \ket{-x,0} $ corresponds to the ground state of the system before the level crossing. It is worth noting that when the Stark coupling $ U $ is zero, these results align with those of the quantum Rabi model~\cite {PhysRevA.86.015803}.

\begin{figure}[t]
	\centering
	\includegraphics[angle=0,width=0.85\linewidth]{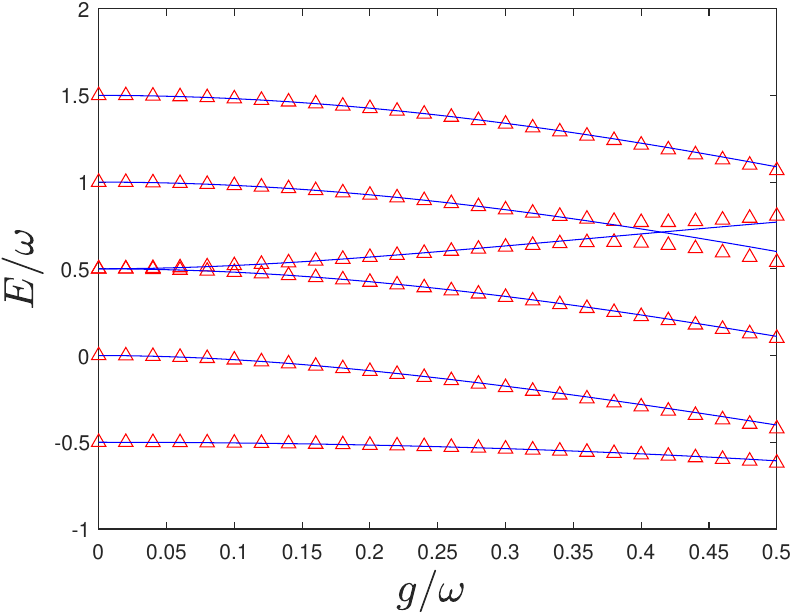}
	\renewcommand{\figurename}{FIG}
	\caption{(Color online) The ground-state energy and the first five excited-state energy of the Rabi-Stark model are plotted as a function of the Rabi coupling $ g $ in units of $\omega$, with $\Delta = \omega$ and $U = \omega$. The red triangle symbols represent the truncated numerical results, and the blue lines depict the results obtained using our method.}
	\label{Fig3}
\end{figure}

\begin{figure}[b]
	\centering
	\includegraphics[angle=0,width=0.95\linewidth]{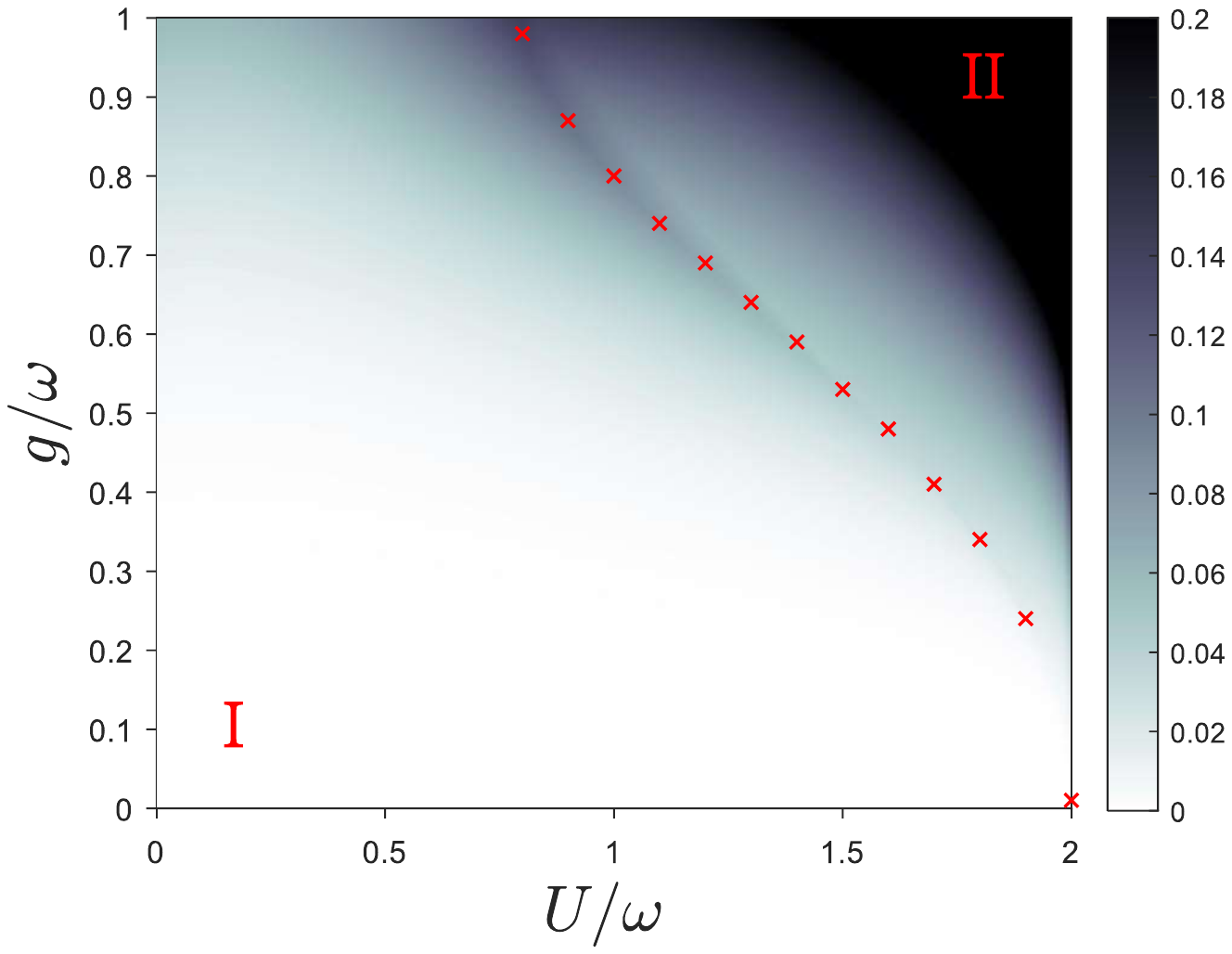}
	\renewcommand{\figurename}{FIG}
	\caption{(Color online) The errors $ \delta E $  of the ground state energy versus the Rabi coupling $ g $ and the Stark coupling $ U $ in units of $ \omega $. The error is calculated as $ \delta E = | E_\text{g} - E_\text{RS} | $, where $ E_\text{g} $ is the ground state energy obtained through our analytical method, and $ E_\text{RS} $ is the ground state energy obtained through truncated numerical simulations. The red cross symbols are obtained by calculating the positions where two energy levels intersect, resulting in a twofold degenerate ground state.}
	\label{FIG5}
\end{figure}

The Hamiltonian of the Rabi-Stark model is now simplified to a two-dimensional form, allowing for the obtainment of its approximate solution with convergence in the ultrastrong coupling regime through this method.\textbf{ Similar to the Jaynes-Cummings model, the energy levels can be labeled by the excitation number, denoted as $ E_{\ket{n,\pm x}} $, implying that this model exhibits near ``superintegrability"~\cite{PhysRevLett.107.100401} for smaller values of the Rabi coupling, within the ultrastrong coupling regime. However, for larger values of the Rabi coupling, an additional ``good quantum number" that uniquely labels each energy level cannot be found, except for parity and energy~\cite{Eckle_2017}, indicating the limitations of our analytical method.} To validate our results, a comparison with numerical simulations will be conducted, and any discrepancies will be clearly observed in the subsequent figures.

In Fig. \ref{Fig3}, we present the energy levels of the Rabi-Stark model, including the ground state energy and the first five excited states, as a function of the Rabi coupling $ g $. Our results exhibit good agreement with the direct numerical simulations, except for a potential loss of accuracy when the Rabi coupling $g$ approaches approximate $0.5\omega$. Furthermore, noticeable discrepancies between the analytical and numerical results are observed in the immediate vicinity of the level crossing or avoided level crossing positions~\cite{Eckle_2017} depicted in Fig. \ref{Fig3}. These results suggest that the relative error in our analytical method is primarily influenced by larger values of the Rabi coupling $g$ and the occurrence of level crossings.


\begin{figure}[tbp]
	\centering
	\includegraphics[angle=0,width=0.95\linewidth]{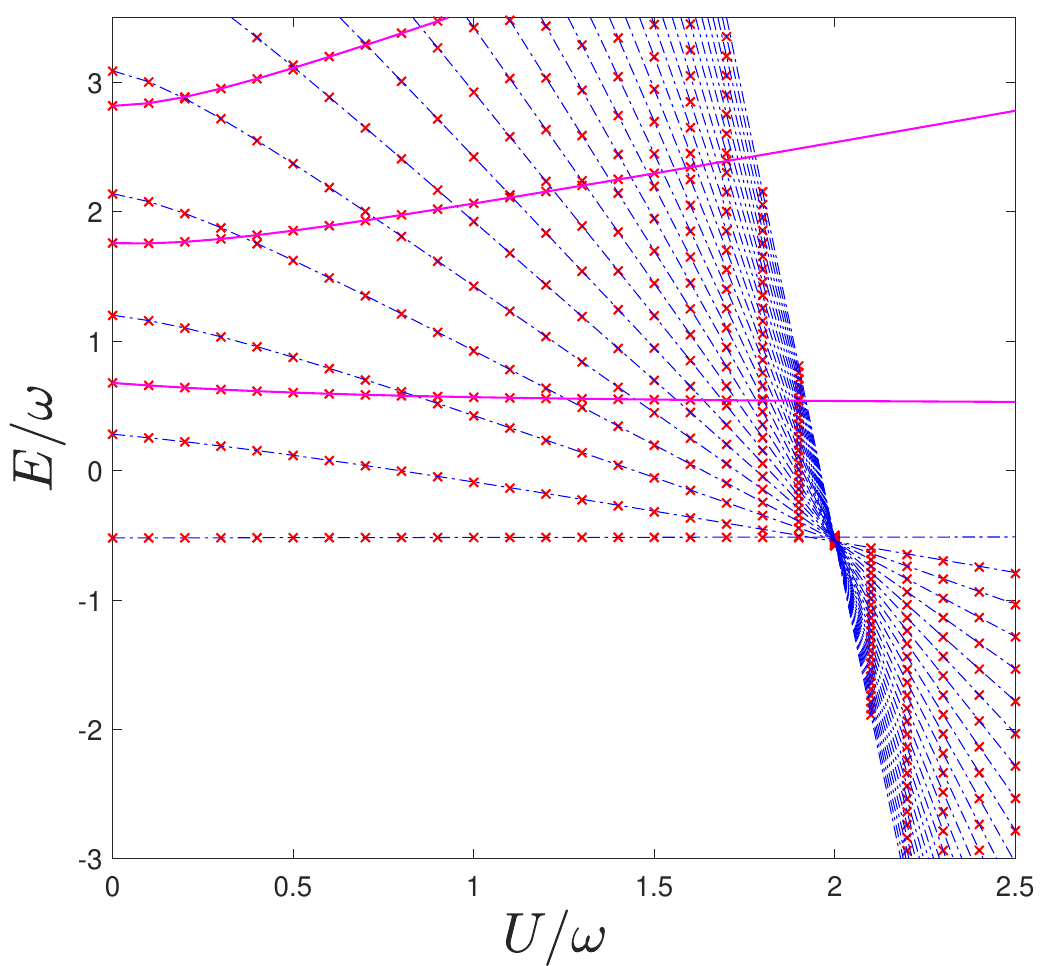}
	\renewcommand{\figurename}{FIG}
	\caption{(Color online) The energy spectrum of the Rabi-Stark model is shown as a function of the Stark coupling $U$ in units of $\omega$, with $\Delta = \omega$ and $g = 0.2\omega$. The positive and negative branches of the energy are represented by solid and dotted lines, respectively. The red cross symbols correspond to the truncated numerical results of the first thirty lowest energy levels displayed in the figure. Both the solid and dotted lines depict the results obtained through our analytical method.}
	\label{Fig4}
\end{figure} 


The validity of our analytical method is further demonstrated by plotting the error of the ground state energy in Fig. \ref{FIG5}. In region I, the analytical solution of the ground state energy is given by Eq. (\ref{gnd-eng}), while in region II, it is determined by taking the minimum eigenvalue of Eq. (\ref{erchenger}). The red cross in Fig. \ref{FIG5} represents the exact position where the ground state energy intersects with the ``first excited state". For a given value of $U$, the error initially increases with an increasing Rabi coupling $g$ before reaching a local maximum at the level crossing. Subsequently, the error slightly decreases before rapidly increasing again as $g$ continues to increase. These observations suggest that large values of $g$, $U$, and the presence of a level crossing are the primary factors leading to inaccuracies in our analytical method. This aligns with the nature of our perturbation-based approach, which inherently faces challenges near level crossings. Overall, Fig. \ref{FIG5} provides insight into the effective range of our analytical method, specifically in region I and the lower left portion of region II. Additionally, it is noteworthy that our method remains valid for smaller values of $g$ (approximately $g \lesssim 0.3\omega$), even in the presence of a level crossing with a large $U$, as illustrated in Fig. \ref{FIG5}.

Fig. \ref{Fig4} depicts the energy spectra obtained using both the truncated numerical method and our analytical approach for the Rabi-Stark model under varying Stark couplings $ U $. Remarkably, our analytical method consistently produces energy spectra that exhibit excellent agreement with the truncated numerical results. Importantly, this accuracy is maintained even when considering additional energy levels in the plot. Thus, our approach proves effective in investigating the Rabi-Stark model for smaller values of $g$ within the ultrastrong coupling regime. Fig. \ref{Fig4} also highlights the steepening slope of the corresponding energy level with respect to $U$, ultimately approaching infinity. This observation signifies that as the value of the Stark coupling $U$ surpasses the critical point $U = 2\omega$, marking a quantum phase transition~\cite {PhysRevA.102.063721}, the energy level $ E_{\ket{n, - x}} $ associated with the ground state no longer exists, and the corresponding value of $ n $ becomes infinite. This phenomenon arises from the unbounded nature of the spectrum in this regime, which lacks a ground state and is counterintuitive in physics.

\begin{figure}[tbp]
	\centering
	\includegraphics[angle=0,width=0.85\linewidth]{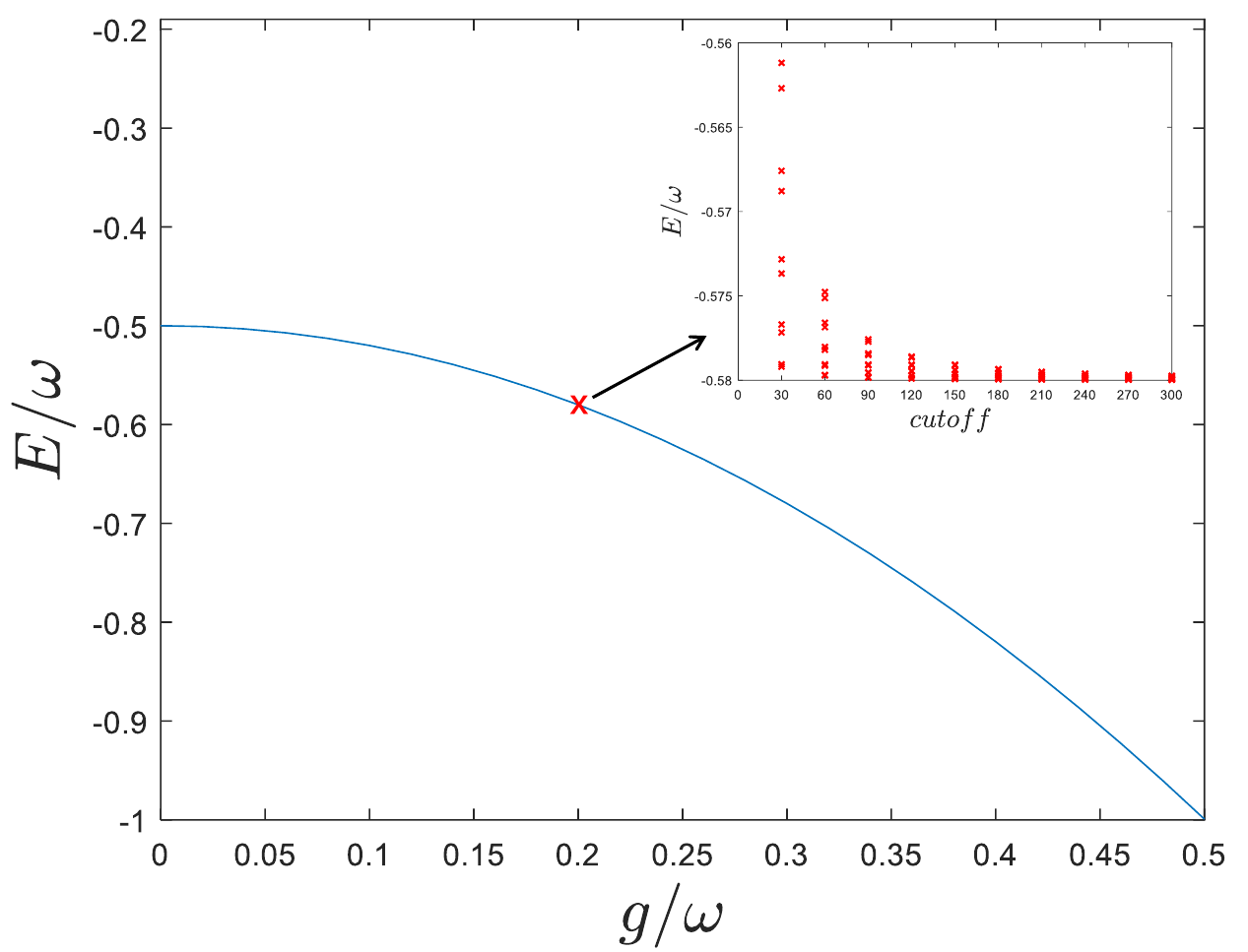}
	\renewcommand{\figurename}{FIG}
	\caption{(Color online) The ground state energy is plotted as a function of the Rabi coupling $ g $ for a fixed Stark coupling $ U = 2\omega $ using the truncated numerical method. Inset: the first ten lowest energy are shown as a function of the number of cutoffs for the Stark coupling $ U = 2\omega $, with $\Delta = \omega$ and $ g = 0.2\omega $.}
	\label{Fig6}
\end{figure}

In Fig. \ref{Fig4}, the energy levels exhibits a notable degeneracy due to the collapse of negative branches when $U = 2\omega$. To further investigate this behavior, we present in Fig. \ref {Fig6} the ground state energy as a function of the Rabi coupling $g$ while maintaining a fixed Stark coupling of $U = 2\omega$, using the truncated numerical method. The inset of Fig. \ref {Fig6} shows the first ten lowest energy levels as the Hamiltonian cutoff is varied, demonstrating that the energy levels collapse to a convergent point. This convergence signifies the presence of a pronounced degeneracy at this specific value of the Stark coupling, indicating the existence of a well-defined ground state energy.


\subsection{Original Model in the CO Limit}

\textbf{The presence of spectrum collapse phenomenon at finite frequency ratios motivates us to analytically investigate its causes and determine whether it disappears in the classical oscillator limit (CO limit), where the quantum phase transition is usually considered. To achieve this, we employ common decoupling methods in the CO limit to derive a clear analytical expression for the excitation energy.} We start by considering the Jaynes-Cummings-like model Hamiltonian given by Eq. (\ref{hamidun}) in the CO limit. \textbf{This limit implies an infinitely large ratio between the qubit frequency and the field frequency, denoted as $\omega \ll \Delta$, resulting in outcomes that are equivalent to those attained in the thermodynamic limit.} Within the ultrastrong coupling regime, which encompasses the range $0.1\omega < g < \omega$, Eq. (\ref{hamidun}) can be simplified as
\begin{equation}\label{zhuanhuandeH}
\begin{split}
	H_{\text{eff}} &= \omega a^{\dagger} a + \frac{\Omega(n)}{2}\tau_{z} + \omega \lambda^2 + 2\lambda g + \frac{U}{2} \tau_{z} a^{\dagger} a \\ 
	&\quad + \Gamma(n)(\tau_{+} a + \tau_{-} a^{\dagger}) ,
\end{split}
\end{equation}
where $ \Omega(n) = (\Delta - 2U \lambda^2 )e^{-2\lambda^2 } \simeq \Delta $, $ \Gamma(n) = g +\lambda(G  - U n - \Delta ) $, $ G = \omega - U T_z/2 $, and the limit condition on $ \lambda $ is transformed to
\begin{equation}
\begin{split}
	\lambda \simeq - \frac{g}{ U n +\Delta + G} .
\end{split}
\end{equation} 
To decouple the interaction induced by the atomic operator, we employ a unitary transformation using the operator $ \hat{B} = \exp[ \Gamma(n) / \Omega(n) (a^{\dagger} \tau_{-} - a \tau_{+})] $. This transformation allows us to express Eq. (\ref{zhuanhuandeH}) as
\begin{equation}\label{SW}
\begin{split}
	H_{\text{eff}} &\simeq \\
	&\omega a^{\dagger} a + \frac{\Delta}{2}\tau_{z} + \omega \lambda^2 + 2\lambda g + \frac{U}{2} \tau_{z} a^{\dagger} a + \frac{\Gamma(n)^2}{\Delta} \tau_{z} a^{\dagger} a  ,
\end{split}
\end{equation}
in the normal phase, our focus lies on the low energy state, which means $ n $ can be considered small. By projecting Eq. (\ref{SW}) to the lower energy level of the two-level system subspace, we get 
\begin{equation}\label{qiguai}
\begin{split}
	H_{\text{eff}} = (\omega - \frac{U}{2} - C) a^{\dagger} a - \frac{\Delta}{2} + \omega \lambda^2 + 2\lambda g  ,
\end{split}
\end{equation}
where $ C = 4 g^2 G^2 / [\Delta (Un + \Delta + G)^2]   \simeq 0 $,  $ E_\text{np} = - \Delta / 2 + \omega \lambda^2 + 2\lambda g \simeq - \Delta / 2 + 2\lambda g $ represents the ground state energy without excitation, consistent with Eq. (\ref{gnd-eng}) in the CO limit. the excitation energy is thus $ \epsilon_\text{np} = \omega - U / 2 - C $. It is important to note that Eq. (\ref{qiguai}) holds valid only under the CO limit and in the ultrastrong coupling regime. Consequently, the final term in $\epsilon_\text{np}$ contributes little, allowing us to rewrite the excitation energy as $\epsilon_\text{np} = \omega - U/2$. When the excitation energy $ \epsilon_\text{np} = 0 $, it indicates that the field mode is macroscopically occupied, signifying the solution of $ U = 2\omega $, This condition serves as the phase boundary between the normal phase and the superradiant phase. 

The above solution can be explained by the Hamiltonian expression in Eq. (\ref{yuanshimao}). By projecting $ \sigma_{z} $ onto the lower energy level of the two-level system subspace with Stark coupling $ U = 2\omega $, the Hamiltonian can be expressed as $ \hat{H}_\text{RS} = - \Delta/2 + g (a^{\dagger} + a)\sigma_{x} $. In the ultrastrong coupling regime, the latter term involving $ g $ is considerably weaker compared to the former term in the CO limit, Consequently, this term $ g (a^{\dagger} + a)\sigma_{x} $ can be disregarded. Thus, the photon occupation number has minimal impact on the overall energy at $ U = 2\omega $, This observation signifies the emergence of high degeneracy at $ U = 2\omega $, which explains the occurrence of spectral collapse. When $U > 2\omega$ and the excitation energy $\epsilon_{np}$ is negative, the energy decreases with an increasing number of photons. In such a scenario, there is no value of $n$ that minimizes the energy, resulting in an unbounded Hamiltonian from below. 

From this analysis, we observe that although the spectrum collapse phenomenon persists in the CO limit, we gain a deeper comprehension of its underlying mechanisms.\textbf{ Notably, these findings align with the behavior exhibited by the Rabi model~\cite{PhysRevLett.115.180404}  when $U = 0$ in the regime of ultrastrong coupling and CO limit, where the Rabi model simplifies to a trivial form. However, our results possess greater physical significance for $ U\neq 0 $, as they provide valuable insights into the causes of the spectral collapse phenomenon.} It is worth noting that the analytical derivation in the so-called superradiant phase, where the ground state is absent, is meaningless and thus omitted in this discussion.


\subsection{Adding Nonlinear Photon-Photon Interaction Term}

From the energy spectra shown in Fig. \ref{Fig4}, it is evident that the quantum Rabi-Stark model experiences spectral collapse and demonstrates an incomplete physical characteristic in the ultrastrong coupling regime when the Stark coupling $U$ surpasses the critical point $U = 2\omega$. To stabilize this system, ensure a bounded Hamiltonian from below, and eliminate the occurrence of the spectral collapse, we introduce a nonlinear term quadratic in $a^\dagger a$, representing photon-photon interaction, to the original Rabi-Stark model Hamiltonian. The modified Hamiltonian is provided in Eq. (\ref{wanzhengmao}).

Utilizing the solution derived in the previous section, we employ a similar approach to solve the present model. The resulting transformed effective Hamiltonian takes the form
\begin{equation}
\begin{split}
	H_\text{E}^{'} &= H_\text{d} + H_\text{nd} + H_{\kappa d} + H_{\kappa nd} \\
	&= \omega a^{\dagger} a + \omega \lambda^2 + 2\lambda g +(\frac{\Delta + U \lambda^2 }{2}) \tau_{z} G_{0}(n) \\
	&+ \frac{U}{2} \lambda \tau_{z} [F_{1}(n) a^{\dagger} a - a F_{1}(n + 1) a^{\dagger} ] + \frac{U}{2} \tau_{z} G_{0} a^{\dagger} a\\
	&+ \kappa [\lambda^4 + \lambda^2 + 4\lambda^2 a^{\dagger} a + a^{\dagger} a a^{\dagger} a] + (\tau_{+} a + \tau_{-} a^{\dagger} )\\
	&\quad \times (\lambda \omega + g + 2 \kappa \lambda^3 + 2\kappa \lambda a^{\dagger} a - \kappa \lambda T_{z} - R_{+}(\lambda)) ,
\end{split}
\end{equation}
where $ \lambda $ satisfies the equation:
\begin{equation}
\begin{split}
	\lambda \omega + g + 2 \kappa \lambda^3 + 2 \kappa \lambda a^{\dagger} a + \kappa \lambda T_{z} + R_{-}(\lambda) = 0,
\end{split}
\end{equation}
Furthermore, by considering the $ 2\times2 $ subspace, similar to Eq. (\ref{erchenger}), the excited state energy can be determined as
\begin{equation}\label{tianjiajuzhen}
\begin{split}
	&H_{11}^{'} = H_{11} + \kappa[\lambda^4 + \lambda^2 + 4\lambda^2 n + n^2 ],\\
	&H_{12}^{'} = H_{12} + 2\kappa \lambda^3 + \kappa \lambda(2 n + 1),\\
	&H_{21}^{'} = H_{21} + 2\kappa \lambda^3 + \kappa \lambda(2 n + 1),\\
	&H_{22}^{'} = H_{22} + \kappa[\lambda^4 + \lambda^2 + 4\lambda^2 (n + 1) + (n + 1)^2 ],
\end{split}
\end{equation}
Moreover, the energy expectation value of $ \ket{-x, 0} $ can be expressed directly as
\begin{equation}\label{EG}
\begin{split}
	E_{0}^{'} &= \omega \lambda^2 + 2 \lambda g \\
	&+ \left( \frac{U \lambda^2 - \Delta}{2} - 2 U \lambda^4  \right) e^{-2\lambda^{2}} + \kappa \lambda^2(1 + \lambda^2 ) .
\end{split}
\end{equation}


\begin{figure}[tbp]
	\centering
	\includegraphics[angle=0,width=0.95\linewidth]{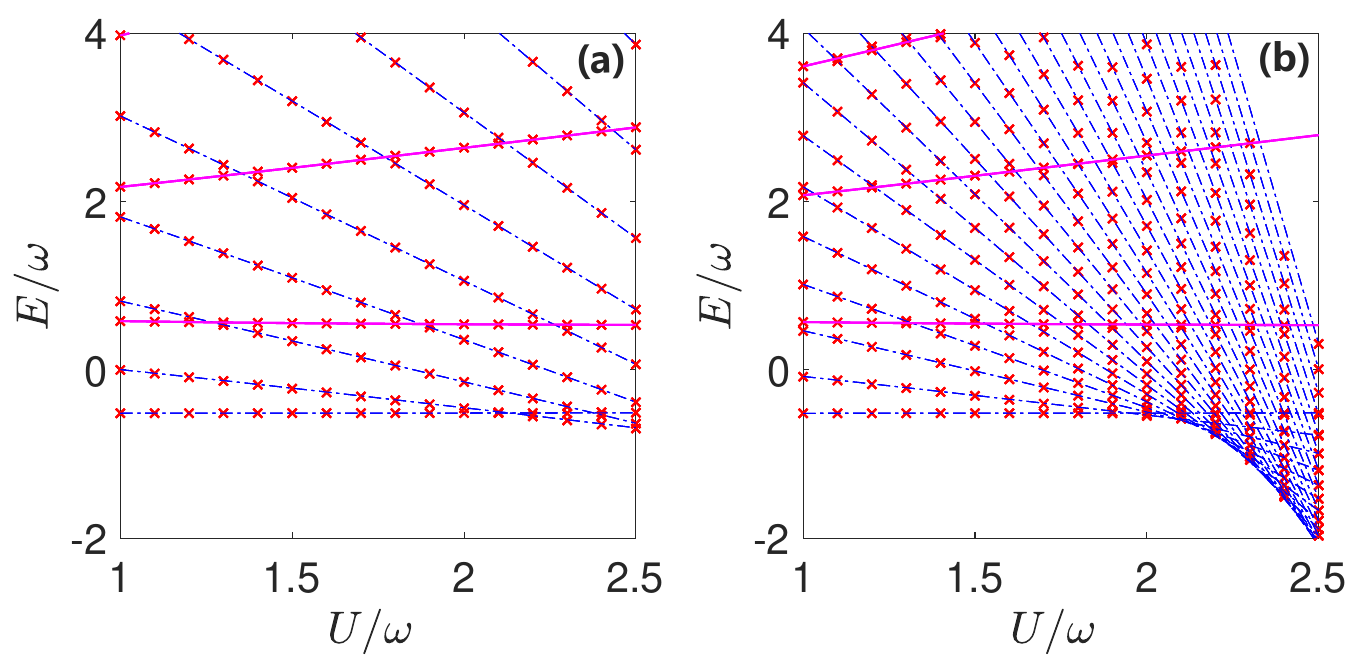}
	\renewcommand{\figurename}{FIG}
	\caption{(Color online) The energy spectrum of the completed Rabi-Stark model is shown as a function of the Stark coupling $ U $ with $\Delta = \omega$ and $ g = 0.2\omega $: (a) $ \kappa = 0.1 \omega$; (b) $ \kappa = 0.01 \omega$. The solid and dashed lines correspond to the positive and negative branches of energy obtained using our method. The red cross symbols represent the numerical results of the first thirty lowest energy levels obtained through the truncated numerical method.}
	\label{Fig7}
\end{figure}
 
Fig. \ref{Fig7} displays the energy spectra of the completed Rabi-Stark model as a function of the Stark coupling $U$. Our analytical results demonstrate good agreement with the truncated numerical results. In contrast to the original Rabi-Stark model shown in Fig. \ref {Fig4}, the lowest energy level no longer exhibits divergence when the value of the Stark coupling $ U $ exceeds the original critical point. This implies the existence of a specific ground state energy  level in the system. The inclusion of a nonlinear photon term ensures the system remains well-defined even for $U > 2\omega$. Notably, the first energy level crossing occurs after the original critical point without the presence of high degeneracy, signifying the elimination of spectral collapse.


\subsection{Completed Model in the CO Limit}

\begin{figure}[t]
	\centering
	\includegraphics[angle=0,width=0.85\linewidth]{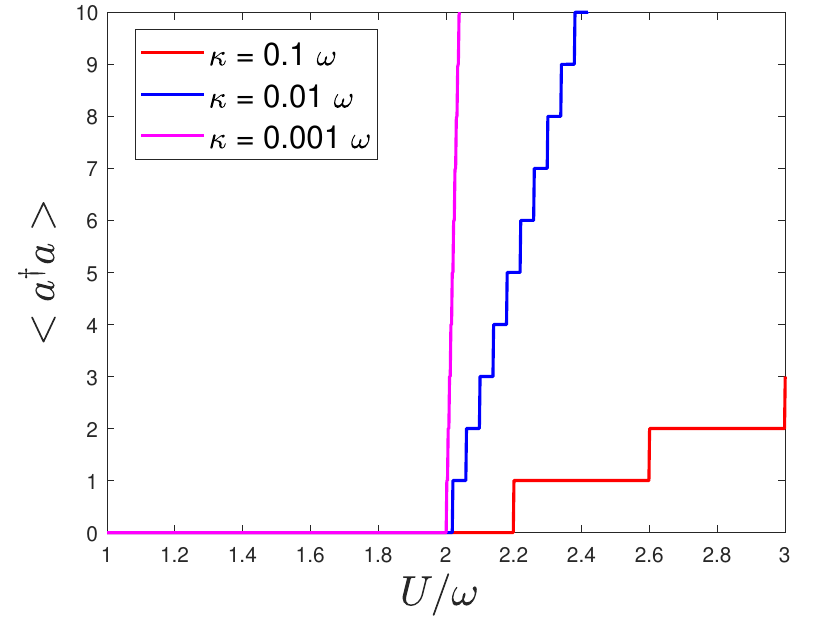}
	\renewcommand{\figurename}{FIG}
	\caption{(Color online) The mean photon number of the ground state in the completed Rabi-Stark model is shown as a function of the Stark coupling $ U $ for different photon coupling $ \kappa $ in units of $ \omega $, with $ g = 0.1\omega$ and $ \Delta = 200 \omega$.}
	\label{Fig8}
\end{figure}

With the successful avoidance of spectrum collapse through the introduction of the nonlinear term, our attention now turns to the behavior of the completed Rabi-Stark model in the classical oscillator (CO) limit and the presence of a possible phase transition.

The mean photon number, as shown in Fig. \ref{Fig8}, exhibits a distinctive ``staircase" pattern as a function of the Stark coupling $U$ in the CO limit, i.e., $\omega \ll \Delta$. In this case, each step of the ``staircase" shows a uniform width, and as the parameter $\kappa$ increases, the first step, representing the intersection point of the ground and first excited energy levels, shifts towards larger values of $U$, contrary to the critical point at $U = 2\omega$ in the original model. Conversely, as $\kappa$ decreases, the step widths in the staircase pattern become narrower, resulting in a steeper staircase, and ultimately diverging to infinity as $\kappa$ approaches zero. This divergence signifies the transition of the completed Rabi-Stark model back to the original Rabi-Stark model, where the phase transition occurs.

\begin{figure}[tbp]
	\centering
	\includegraphics[angle=0,width=0.95\linewidth]{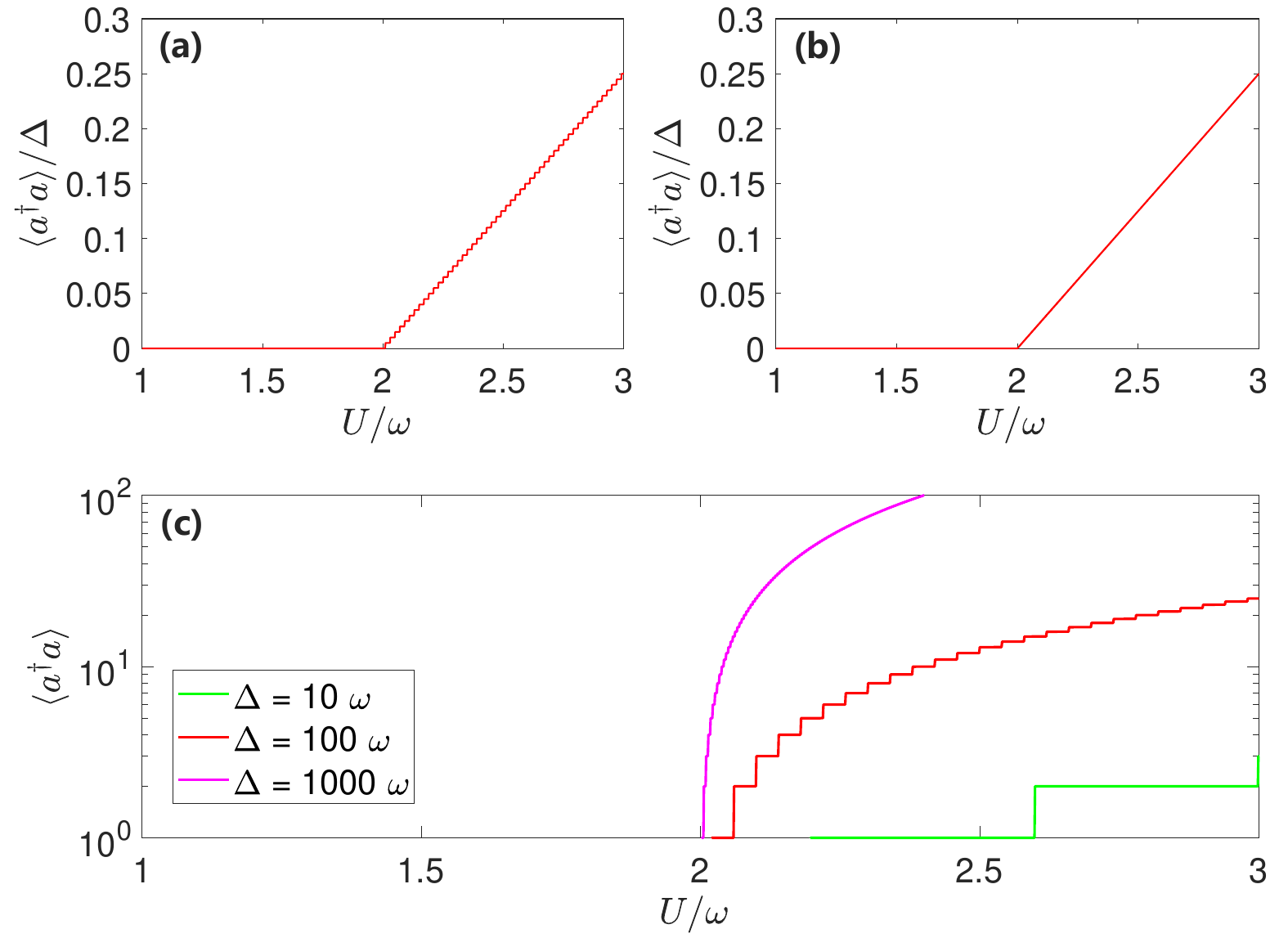}
	\renewcommand{\figurename}{FIG} 
	\caption{(Color online) The renormalized mean photon number of the ground state in the completed Rabi-Stark model is shown as a function of the Stark coupling $ U $ in units of $ \omega $, with $ g = 0.1\omega$, $ \kappa = 1 / \Delta $ and (a) $ \Delta = 200 \omega $, (b) $ \Delta = 1000 \omega $. (c) The mean photon number of the ground state in the completed Rabi-Stark model versus the Stark coupling $ U $ for different $ \Delta $ in units of $ \omega $, with $ g = 0.1\omega$, $ \kappa = 1 / \Delta $. The behavior of the mean photon number becomes sharper near the critical point as the ratio $ \Delta/\omega $ increases.}
	\label{Fig9}
\end{figure}

\textbf{Considering both $\Delta$ needs to be large enough in the CO limit and $\kappa$ needs to be small enough to preserve some physical features of original model such as phase transition, we set $\kappa = 1/\Delta$ and plot the renormalized mean photon number, $a^\dagger a / \Delta$ in Fig. \ref{Fig9} as a function of $U$ while keep $\Delta/\omega \to \infty$. Interestingly, it is observed that the latter part of the mean photon number function displays a linear behavior with a fixed slope.}

In Fig. \ref{Fig9}(c), employing a logarithmic y-axis for clarity, it becomes evident that as $\Delta$ increases, the mean photon number of the ground state exhibits a sharper change in the vicinity of the critical point when $\kappa = 1/\Delta$. The mean photon number of the ground state serves as an order parameter and this behavior bears resemblance to the second-order transition observed in the Jaynes-Cummings model within the classical oscillator limit (CO limit)~\cite {PhysRevLett.117.123602}.

Next, we analyze the features shown above using analytical methods. In the ultrastrong coupling regime, the completed Rabi-Stark model can be approximately reduced to a $2 \times 2$ Hilbert subspace as shown in Eq. (\ref{tianjiajuzhen}). In the CO limit, where the parameter $\Delta/\omega$ becomes infinitely large, the constraint condition for the parameter $\lambda$ can be approximated as,
\begin{equation}
\begin{split}
	\lambda \simeq - \frac{g}{ U n + \Delta + G + 2\kappa n - \kappa T_z } ,
\end{split}
\end{equation}

By considering terms up to first-order precision in the matrix elements of the Hamiltonian within the $2 \times 2$ subspace and neglecting the term that contributes insignificantly compared to other terms in the expression for the eigenenergies, the lower eigenenergy of this energy matrix can be written as
\begin{equation}\label{20}
\begin{split}
	E_{\ket{n, -x}} =
- \frac{ 1 }{ 2 }n U - \frac{\Delta}{2} + n(n\kappa + \omega) + 2\lambda g .
\end{split}
\end{equation}
where $n = 1, 2, \ldots$. Noting that the Eq. (\ref{EG}) is consistent with this equation by taking $n = 0$ in CO limit, the negative branches of the spectrum can be approximated by this unified equation with  $n = 0, 1, 2, \ldots$.

From this equation, we can obtain that the level crossing position as 
\begin{equation}\label{Udehaojieguo}
	\begin{split}
		U = 2\omega + 2\kappa + 4n \kappa, \qquad  n = 0,1,2\cdots .
	\end{split}
\end{equation}
When n = 0, the first intersection position is $U = 2\omega + 2\kappa$, which illustrates the influence on the phase boundary of the adding nonlinear photon term. Moreover, the distance between neighboring intersection positions becomes $U = 4\kappa$, which explains the equal step width in Fig. \ref{Fig8}. 

The linear relationship between the renormalized mean photon number and the Stark coupling $U$ can also be further confirmed by considering Eq. (\ref{Udehaojieguo}) in the form $U = 2\omega + 2\kappa + 4n/\Delta$ under the condition $\kappa = 1/\Delta$, which yields a slope of $l = 1/4$. Additionally, a more detailed analysis using a geometric approach in Appendix B leads to the same result.

In summary, within the ultrastrong coupling regime and for small photon coupling $\kappa$, as the Stark coupling $U$ exceeds a certain threshold, the mean photon number of the ground state transitions from $0$ to $1$ instead of diverging to infinity. This transition is characterized by a ``staircase" pattern in the function of the mean photon number. As the photon coupling $\kappa$ increases, the step width in the staircase pattern, which analytically written at $4\kappa$, widens. This observation indicates that larger values of photon coupling $\kappa$ have a significant influence on the behavior of the mean photon number for Stark coupling values beyond the critical point, impeding photon transitions between energy levels. Furthermore, the completed Rabi-Stark model exhibits a phase boundary shifted by $2\kappa$ compared to the original model, resulting in a higher critical point. Notably, the mean photon number demonstrates a linear relationship with the Stark coupling $U$, featuring a fixed slope of $l=1/4$ as the photon coupling $\kappa$ approaches $1/\Delta$.

\section{CONCLUSIONS}

In this study, we present an analytical approach to approximate the quantum Rabi-Stark model in the ultrastrong coupling regime by transforming it into a solvable Jaynes-Cummings-like model. We validate the effectiveness of our analytical method by comparing our results with direct numerical calculations. By analyzing the energy spectra shown in Fig. \ref{Fig4}, we observe the occurrence of the spectral collapse phenomenon at the critical point $U = 2\omega$, which marks the boundary of the superradiant phase transition. When the Stark coupling $U > 2\omega$, the system lacks a ground state, resulting in an unbounded energy. To address this instability, we introduce a nonlinear photon term $\kappa(a^\dagger a)^2$, representing photon-photon interaction, to the Hamiltonian. Remarkably, this additional term stabilizes the model and eliminates the spectral collapse phenomenon, allowing the system to possess a well-defined ground state and to be completed.

Besides, we investigate both the completed Rabi-Stark model and the original Rabi-Stark model in the classical oscillator (CO) limit. By neglecting higher-order terms with negligible contributions to the energy, we derive a simplified analytical expression for the energy spectra. We provide a theoretical explanation for the macroscopic occupation of the ground state and the occurrence of spectral collapse in the original model. For the completed Rabi-Stark model in the CO limit, we discover that the energy spectrum of the negative branches exhibits a linear relationship, and all energy levels can be described by a unified expression [Eq. (\ref{20})]. The mean photon number of the ground state displays a ``staircase" pattern with a constant step width of $4\kappa$, as demonstrated analytically.\textbf{ The photon coupling $\kappa$ influences photon transitions between energy levels.} Specifically, when $\kappa$ approaches $1/\Delta$, the mean photon number of the ground state exhibits a fixed slope of $l = 1/4$, which distinguishes it from the behavior of the original Rabi-Stark model. \textbf{Moreover, the mean photon number undergoes a second-order transition, reminiscent of the behavior observed in the Jaynes-Cummings model within the CO limit.} We also investigate the impact of the additional photon-photon interaction term on the quantum phase transition. \textbf{The phase boundary is shifted by $2\kappa$, leading to a new critical point at $U = 2\omega + 2\kappa$.} This term transforms the superradiant phase into a ``completed" phase throughout the coupling regime, in contrast to the unstable superradiant phase of the original Rabi-Stark model. Our findings shed light on the study of the unbounded-from-below superradiant phase transition caused by the spectral collapse phenomenon.

\section*{ACKNOWLEDGMENTS}
This work was supported by NSFC under grants Nos. 12074027. 

\appendix

\section{Simplification of the Rabi-Stark Model}

We apply a unitary transformation using the operator $ \exp[\lambda\sigma_z(a^{\dagger} - a)] $ to the Hamiltonian[Eq.(\ref{xuanzhuanmao})], where $ \lambda $ is a parameter constrained by the following condition. Following this transformation, the effective Hamiltonian can be divided into four distinct parts, 
\begin{eqnarray}\label{yaozhengnmao}
	\hat{H}_\text{E} = \hat{H}_1+\hat{H}_2+\hat{H}_3+\hat{H}_4,
\end{eqnarray}
where 
\begin{equation}
\begin{split}
	&\hat{H}_1 = \omega a^{\dagger} a-\lambda\omega\sigma_z\left(a^{\dagger} +a\right)+\omega\lambda^2,\\
	&\hat{H}_2 = \frac{\Delta}{2}\{\sigma_x \cosh[2\lambda(a^\dagger-a)]+i \sigma_y\sinh[2\lambda(a^\dagger-a)]\},\\
	&\hat{H}_3 = -g[\sigma_z\left(a^\dagger+a\right)-2\lambda],\\
	&\hat{H}_4 = 
	\frac{U}{2}\cosh[2\lambda(a^\dagger-a)]
	[   \sigma_x(\lambda^2+a^\dagger a)   +  i\sigma_y \lambda (a^\dagger+a) ]\\
	&+
	\frac{U}{2}\sinh[2\lambda(a^\dagger-a)]
	[\lambda\sigma_x(a^\dagger+a)   +  i\sigma_y(\lambda^2+a^\dagger a)].
\end{split}
\end{equation}
Here the terms $\cosh[2\lambda(a^\dag-a)]$ and $\sinh[2\lambda(a^\dag-a)]$ can be expanded by the expansion formula of $\cosh(x)$ and $\sinh(x)$. These two terms can be written as
\begin{equation}
	\begin{split}
	\cosh[2\lambda(a^\dagger-a)]&=G_0(N)+G_1(N)a^{\dagger 2} + a^2 G_1(N)+\cdots,\\
	\sinh[2\lambda(a^\dagger-a)]&=F_1(N)a^\dagger-a F_1(N) \\
	&+F_2(N)a^{\dagger 2}-a^2F_2(N)+\cdots,
	\end{split}
\end{equation}
where $G_i(N) (i = 0,1,2,...)$ and $F_j(N) (j = 1,2,...)$ are associated to the parameter $\lambda$ and the photon number $n$. Therefore, other terms in [Eq.(\ref{yaozhengnmao})] can be simplified by
\begin{flalign}
	\begin{aligned}
	&\sinh{[2\lambda(a^\dagger-a)]} (a^\dagger+a)=\\
	&F_1(N)a^{\dagger 2}+F_1(N)a^\dagger a
	-aF_1(N)a^\dagger-aF_1(N)a+\cdots,\\
    &\cosh{[2\lambda(a^\dagger-a)]} (a^\dagger+a)=G_0(N)a^\dagger+G_0(N)a+\cdots,\\
    &\sinh{[2\lambda(a^\dagger-a)]} a^\dagger a=[F_1(N)a^\dagger -a F_1(N)]a^\dagger a+\cdots,\\
    &\cosh{[2\lambda(a^\dagger-a)]} a^\dagger a=G_0(N)a^\dagger a + G_1(N) a^{\dagger 3} a +\cdots.
    \end{aligned}
\end{flalign}
The high-order terms of $a$ and $a^{\dagger} $ are neglected since they primarily impact the far off-diagonal elements of the Hamiltonian. These terms make minimal contributions to the total energy, indicating their relevance within perturbation theory. As a result, the effective Hamiltonian can be simplified to
\begin{equation}\label{bianhuanhouzongh}
	\begin{split}
	&\hat{H}_\text{E} = \hat{H}_1+\hat{H}_2+\hat{H}_3+\hat{H}_4\\
	&= \omega a^\dagger a-(\lambda\omega+g)\sigma_z\left(a^\dagger+a\right)+\omega\lambda^2 + 2\lambda g\\
	&+(\frac{\Delta}{2}+\frac{U}{2}\lambda^2)\left\{\sigma_x G_0(n) + i\sigma_y\left[F_1(n) a^\dagger-a F_1(n)\right]\right\}\\
	&+\frac{U}{2}\sigma_x\left[\lambda F_1(n) a^\dagger a-\lambda aF_1(n) a^\dagger  +   G_0(n)a^\dagger a\right]\\
	&+\frac{i U}{2}\sigma_y\left\{  [F_1(n)a^\dagger-aF_1(n)]a^\dagger a  +   \lambda G_0(n)(a^\dagger+ a) \right\},\\
    \end{split}
\end{equation}
where $G_0(n)$ and $F_1(n)$ can be calculated directly, with $ G_0(n) =\bra{n}G_0(N)\ket{n} = \bra{n}\cosh[2\lambda(a^\dagger-a)]\ket{n} = $
$ L_n(4\lambda^2) \exp(-2\lambda^2) $, and $ F_1(n) = \bra{n+1}\sinh[2\lambda(a^\dagger-a)]\ket{n} = 2\lambda L_n^1(4\lambda^2) \exp(-2\lambda^2) /(n+1) $.
In the expression above $L_n(x)$ is the Laguerre polynomial, and $L_n^1(x)$ is the associated Laguerre polynomial with superscript 1.

After performing a transformation of the Hamiltonian into the representation defined by $\sigma_x$ (i.e. $\sigma_x\ket{\pm x} = \pm\ket{\pm x}$), where $ \sigma_x = \tau_z,\quad \sigma_y = -i\left(\tau_+-\tau_-\right) ,\quad \sigma_z = -\left(\tau_+ + \tau_-\right) $,
the effective Hamiltonian can be simplified as the sum of  diagonal and non-diagonal terms, denoted by $H_\text{d}$ and $H_\text{nd}$, respectively.
\begin{equation}
	\begin{split}
	&H_\text{d}=\omega a^\dagger a + \omega\lambda^2 + 2\lambda g+(\frac{\Delta}{2}+\frac{U}{2}\lambda^2)\tau_zG_0\left(n\right)\\
	&+ \frac{U}{2}\lambda\tau_z[F_1(n)a^\dagger a - aF_1(n+1)a^\dagger] + \frac{U}{2}\tau_z G_0(n) a^\dagger a,\\
	&H_\text{nd}
	= \left(\tau_+a+\tau_-a^\dagger\right)\left(\lambda\omega+g-R_+(\lambda)\right)\\
	 & \qquad\qquad\quad +(\tau_+a^\dagger+\tau_-a)(\lambda\omega+g+R_-(\lambda)),
	\end{split}
\end{equation}
with
\begin{equation}
	\begin{split}
     &R_+\left(\lambda\right) = \frac{1}{2}F_1(n)\left(\Delta + U\lambda^2 + Un\right) + \frac{U}{2}G_0(n)\lambda T_z, \\
     &R_-\left(\lambda\right) = \frac{1}{2}F_1(n)\left(\Delta + U\lambda^2 + Un\right) - \frac{U}{2}G_0(n)\lambda T_z.
	\end{split}
\end{equation}
In the nondiagonal term, $ T_z $ takes values of $ \pm 1 $ corresponding to the different energy levels of the spin operator associated with $ \tau_{z} $. The parameter $\lambda$ is chosen to satisfy 
\begin{equation}\label{tiaojian}
    \lambda\omega + g - \frac{U}{2}G_0\left(n\right)\lambda T_z + \frac{F_1\left(n\right)}{2}\left(\Delta + U \lambda^2 + U n\right) = 0.
\end{equation}
The resulting effective Hamiltonian takes the form of a  Jaynes-Cummings-like model, exhibiting diagonal elements except for its final term in the Hamiltonian
\begin{align}\label{A9}
	\notag
	H_\text{E} &= \left(\omega+\frac{U}{2}\tau_zG_0(n) \right) a^\dagger a +  \left(\frac{\Delta}{2} + \frac{U}{2}\lambda^2\right)\tau_zG_0(n)\\\notag
    	 &+ (2g+\omega\lambda )\lambda + \frac{U}{2}\lambda\tau_z\left[F_1(n)a^\dagger a - aF_1(n+1)a^\dagger\right]\\
	     &+ \left(\tau_+a + \tau_-a^\dagger\right)\left[\lambda\omega + g - R_+(\lambda)\right].
\end{align}


In the experimental setup where the Rabi coupling $ g $ is relatively small ($g < 0.5 \omega $) in the ultrastrong coupling regime, the parameter $\lambda$ is also small ~\cite {PhysRevA.86.015803}. The Laguerre polynomial and the associated Laguerre polynomial can be expanded up to the zero-order term, $L_n(4\lambda^2) \simeq 1$ and $L_n^1(4\lambda^2) \simeq n + 1$,respectively, leading to an approximate analytical solution
\begin{equation}
	\begin{split}
	&\lambda \simeq \\ 
	&-\frac{g}{\omega+(\mathrm{\Delta} \pm \frac{U}{2}+Un)\exp{\left[-2\left(\frac{g}{\omega+\mathrm{\Delta} \pm U/2+Un}\right)^2\right]}}.
	\end{split}
\end{equation}

From this Hamiltonian expression in Eq. (\ref{A9}) mentioned above, the energy spectrum can be obtained within the subspaces $\{\ket{+x,n}, \ket{-x,n+1}\}$, $ n = 0,1,2\cdots  $, 
\begin{equation}
	H_\text{E} = 
	\left( \begin{array}{c c c}
		H_{11} & H_{12}\\
		H_{21} & H_{22}
	\end{array}
	\right),
\end{equation}
where
\begin{align}
	\notag
	&H_{11} = n \omega + 2 g \lambda +\lambda ^2 \omega + e^{-2 \lambda ^2} L_n\left(4 \lambda ^2\right) \left(\frac{\Delta + n U + \lambda ^2 U}{2}\right) + \lambda ^2 U e^{-2 \lambda ^2} \left[ \frac{ L_{n+1}^1\left(4 \lambda ^2\right)}{n+2} - \frac{L_n^1\left(4 \lambda ^2\right)}{n+1}\right]  , \\
	\notag
	&H_{12} = \sqrt{n+1} \left[g + \lambda \omega - \frac{1}{2} \lambda  U e^{-2 \lambda ^2} L_n\left(4 \lambda ^2\right)-\frac{ \lambda  e^{-2 \lambda ^2} L_n^1\left(4 \lambda ^2\right) \left(\Delta + n U + \lambda ^2 U\right)}{n+1} \right] , \\\notag
	&H_{21} = \sqrt{n+1} \left[ g+\lambda \omega + \frac{1}{2} \lambda  U e^{-2 \lambda ^2} L_{n+1}\left(4 \lambda ^2\right)-\frac{ \lambda  e^{-2 \lambda ^2} L_{n+1}^1\left(4 \lambda ^2\right) \left(\Delta + (n+1) U+\lambda ^2 U\right)}{n+2} \right] ,\\
	\notag
	&H_{22} = (n+1) \omega + 2 g \lambda +\lambda ^2 \omega -e^{-2 \lambda ^2} \left\{ L_{n+1}(4 \lambda ^2) \frac{\Delta + (n+1+ \lambda ^2 ) U}{2} + U \lambda ^2 \left[\frac{ L_{n+2}^1\left(4 \lambda ^2\right)}{n+3} - \frac{L_{n+1}^1\left(4 \lambda ^2\right)}{n+2}\right] \right\} .
\end{align}

Since the Hamiltonian [Eq.(\ref{A9})] takes the Jaynes-Cummings form, the energy expectation value of $ \ket{-x, 0} $ can be directly obtained as
\begin{align}
	\notag
	E_\text{0} &=\bra{-x,0} H_E \ket{-x,0}\\
	&=\omega\lambda^2+2\lambda g - \frac{\Delta - U \lambda^2+4U\lambda^4}{2}e^{-2\lambda^2},
\end{align}
the state $ \ket{-x,0} $ corresponds to the ground state of the system before the level crossing. For zero value of the Stark coupling $ U $, the results are consistent with those of the quantum Rabi model.


\section{Geometric Method for the Fixed Slope}
We demonstrate an alternative geometric method to determine the slope in Fig. \ref{Fig9}. Firstly, we determine the step width as $ 4\kappa $,which is obtained from Eq. (\ref{Udehaojieguo}). Next, we evaluate the mean photon number for the negative branches of interest as follows:
\begin{equation}
\begin{split}
	\left\langle a^{\dagger} a \right\rangle _{\ket{0, -}} = \bra{ \varphi_{0} } a^{\dagger} a \ket{ \varphi_{0} } = \lambda^2 \simeq 0 ,
\end{split}
\end{equation}
where $ \ket{ \varphi_{0} } = e^{i \pi \sigma_{y} /4} e^{-\lambda \sigma_{z} (a^{\dagger} - a)} \ket{ 0, - x } $. For $ n \geq 1 $,
\begin{equation}
\begin{split}
	\left\langle a^{\dagger} a \right\rangle _{\ket{n, -}} &= (n + \lambda^2 ) C_1^2 + (n + 1 + \lambda^2 )C_2^2\\
	&+ 2\lambda \sqrt{n + 1} C_{1} C_{2} ,
\end{split}
\end{equation}
where the values of $ C_{1}$ and $ C_{2} $ are determined by the eigenvector. Therefore, the step height can be expressed as 
\begin{equation}
\begin{split}
 \frac{ \left\langle a^{\dagger} a \right\rangle _{\ket{n + 1, -}} - \left\langle a^{\dagger} a \right\rangle _{\ket{n, -}}}{\Delta} = \frac{1}{\Delta} ,
\end{split}
\end{equation}  
and the slope $ l $ in Fig. \ref{Fig9} can be obtained as
\begin{equation}
\begin{split}
	l = \frac{1/\Delta}{4\kappa} = \frac{1}{4}.  
\end{split}
\end{equation}

\vspace{16em}



\bibliography{Primary_manuscript}

\begin{thebibliography}{36}%
\makeatletter
\providecommand \@ifxundefined [1]{%
 \@ifx{#1\undefined}
}%
\providecommand \@ifnum [1]{%
 \ifnum #1\expandafter \@firstoftwo
 \else \expandafter \@secondoftwo
 \fi
}%
\providecommand \@ifx [1]{%
 \ifx #1\expandafter \@firstoftwo
 \else \expandafter \@secondoftwo
 \fi
}%
\providecommand \natexlab [1]{#1}%
\providecommand \enquote  [1]{``#1''}%
\providecommand \bibnamefont  [1]{#1}%
\providecommand \bibfnamefont [1]{#1}%
\providecommand \citenamefont [1]{#1}%
\providecommand \href@noop [0]{\@secondoftwo}%
\providecommand \href [0]{\begingroup \@sanitize@url \@href}%
\providecommand \@href[1]{\@@startlink{#1}\@@href}%
\providecommand \@@href[1]{\endgroup#1\@@endlink}%
\providecommand \@sanitize@url [0]{\catcode `\\12\catcode `\$12\catcode
  `\&12\catcode `\#12\catcode `\^12\catcode `\_12\catcode `\%12\relax}%
\providecommand \@@startlink[1]{}%
\providecommand \@@endlink[0]{}%
\providecommand \url  [0]{\begingroup\@sanitize@url \@url }%
\providecommand \@url [1]{\endgroup\@href {#1}{\urlprefix }}%
\providecommand \urlprefix  [0]{URL }%
\providecommand \Eprint [0]{\href }%
\providecommand \doibase [0]{http://dx.doi.org/}%
\providecommand \selectlanguage [0]{\@gobble}%
\providecommand \bibinfo  [0]{\@secondoftwo}%
\providecommand \bibfield  [0]{\@secondoftwo}%
\providecommand \translation [1]{[#1]}%
\providecommand \BibitemOpen [0]{}%
\providecommand \bibitemStop [0]{}%
\providecommand \bibitemNoStop [0]{.\EOS\space}%
\providecommand \EOS [0]{\spacefactor3000\relax}%
\providecommand \BibitemShut  [1]{\csname bibitem#1\endcsname}%
\let\auto@bib@innerbib\@empty
\bibitem [{\citenamefont {Rabi}(1936)}]{PhysRev.49.324}%
  \BibitemOpen
  \bibfield  {author} {\bibinfo {author} {\bibfnamefont {I.~I.}\ \bibnamefont
  {Rabi}},\ }\bibfield  {title} {\enquote {\bibinfo {title} {On the process of
  space quantization},}\ }\href {\doibase 10.1103/PhysRev.49.324} {\bibfield
  {journal} {\bibinfo  {journal} {Phys. Rev.}\ }\textbf {\bibinfo {volume}
  {49}},\ \bibinfo {pages} {324--328} (\bibinfo {year} {1936})}\BibitemShut
  {NoStop}%
\bibitem [{\citenamefont {Rabi}(1937)}]{PhysRev.51.652}%
  \BibitemOpen
  \bibfield  {author} {\bibinfo {author} {\bibfnamefont {I.~I.}\ \bibnamefont
  {Rabi}},\ }\bibfield  {title} {\enquote {\bibinfo {title} {Space quantization
  in a gyrating magnetic field},}\ }\href {\doibase 10.1103/PhysRev.51.652}
  {\bibfield  {journal} {\bibinfo  {journal} {Phys. Rev.}\ }\textbf {\bibinfo
  {volume} {51}},\ \bibinfo {pages} {652--654} (\bibinfo {year}
  {1937})}\BibitemShut {NoStop}%
\bibitem [{\citenamefont {Lv}\ \emph {et~al.}(2018)\citenamefont {Lv},
  \citenamefont {An}, \citenamefont {Liu}, \citenamefont {Zhang}, \citenamefont
  {Pedernales}, \citenamefont {Lamata}, \citenamefont {Solano},\ and\
  \citenamefont {Kim}}]{PhysRevX.8.021027}%
  \BibitemOpen
  \bibfield  {author} {\bibinfo {author} {\bibfnamefont {Dingshun}\
  \bibnamefont {Lv}}, \bibinfo {author} {\bibfnamefont {Shuoming}\ \bibnamefont
  {An}}, \bibinfo {author} {\bibfnamefont {Zhenyu}\ \bibnamefont {Liu}},
  \bibinfo {author} {\bibfnamefont {Jing-Ning}\ \bibnamefont {Zhang}}, \bibinfo
  {author} {\bibfnamefont {Julen~S.}\ \bibnamefont {Pedernales}}, \bibinfo
  {author} {\bibfnamefont {Lucas}\ \bibnamefont {Lamata}}, \bibinfo {author}
  {\bibfnamefont {Enrique}\ \bibnamefont {Solano}}, \ and\ \bibinfo {author}
  {\bibfnamefont {Kihwan}\ \bibnamefont {Kim}},\ }\bibfield  {title} {\enquote
  {\bibinfo {title} {Quantum simulation of the quantum rabi model in a trapped
  ion},}\ }\href {\doibase 10.1103/PhysRevX.8.021027} {\bibfield  {journal}
  {\bibinfo  {journal} {Phys. Rev. X}\ }\textbf {\bibinfo {volume} {8}},\
  \bibinfo {pages} {021027} (\bibinfo {year} {2018})}\BibitemShut {NoStop}%
\bibitem [{\citenamefont {Schneeweiss}\ \emph {et~al.}(2018)\citenamefont
  {Schneeweiss}, \citenamefont {Dareau},\ and\ \citenamefont
  {Sayrin}}]{PhysRevA.98.021801}%
  \BibitemOpen
  \bibfield  {author} {\bibinfo {author} {\bibfnamefont {P.}~\bibnamefont
  {Schneeweiss}}, \bibinfo {author} {\bibfnamefont {A.}~\bibnamefont {Dareau}},
  \ and\ \bibinfo {author} {\bibfnamefont {C.}~\bibnamefont {Sayrin}},\
  }\bibfield  {title} {\enquote {\bibinfo {title} {Cold-atom-based
  implementation of the quantum rabi model},}\ }\href {\doibase
  10.1103/PhysRevA.98.021801} {\bibfield  {journal} {\bibinfo  {journal} {Phys.
  Rev. A}\ }\textbf {\bibinfo {volume} {98}},\ \bibinfo {pages} {021801(R)}
  (\bibinfo {year} {2018})}\BibitemShut {NoStop}%
\bibitem [{\citenamefont {Haroche}\ and\ \citenamefont
  {Raimond}(2006)}]{10.1093}%
  \BibitemOpen
  \bibfield  {author} {\bibinfo {author} {\bibfnamefont {Serge}\ \bibnamefont
  {Haroche}}\ and\ \bibinfo {author} {\bibfnamefont {Jean-Michel}\ \bibnamefont
  {Raimond}},\ }\href {\doibase 10.1093/acprof:oso/9780198509141.001.0001}
  {\emph {\bibinfo {title} {Exploring the Quantum: Atoms, Cavities, and
  Photons}}}\ (\bibinfo  {publisher} {Oxford University Press},\ \bibinfo
  {year} {2006})\BibitemShut {NoStop}%
\bibitem [{\citenamefont {Jaynes}\ and\ \citenamefont
  {Cummings}(1963)}]{1443594}%
  \BibitemOpen
  \bibfield  {author} {\bibinfo {author} {\bibfnamefont {E.T.}\ \bibnamefont
  {Jaynes}}\ and\ \bibinfo {author} {\bibfnamefont {F.W.}\ \bibnamefont
  {Cummings}},\ }\bibfield  {title} {\enquote {\bibinfo {title} {Comparison of
  quantum and semiclassical radiation theories with application to the beam
  maser},}\ }\href {\doibase 10.1109/PROC.1963.1664} {\bibfield  {journal}
  {\bibinfo  {journal} {Proceedings of the IEEE}\ }\textbf {\bibinfo {volume}
  {51}},\ \bibinfo {pages} {89--109} (\bibinfo {year} {1963})}\BibitemShut
  {NoStop}%
\bibitem [{\citenamefont {Yu}\ \emph {et~al.}(2012)\citenamefont {Yu},
  \citenamefont {Zhu}, \citenamefont {Liang}, \citenamefont {Chen},\ and\
  \citenamefont {Jia}}]{PhysRevA.86.015803}%
  \BibitemOpen
  \bibfield  {author} {\bibinfo {author} {\bibfnamefont {Lixian}\ \bibnamefont
  {Yu}}, \bibinfo {author} {\bibfnamefont {Shiqun}\ \bibnamefont {Zhu}},
  \bibinfo {author} {\bibfnamefont {Qifeng}\ \bibnamefont {Liang}}, \bibinfo
  {author} {\bibfnamefont {Gang}\ \bibnamefont {Chen}}, \ and\ \bibinfo
  {author} {\bibfnamefont {Suotang}\ \bibnamefont {Jia}},\ }\bibfield  {title}
  {\enquote {\bibinfo {title} {Analytical solutions for the rabi model},}\
  }\href {\doibase 10.1103/PhysRevA.86.015803} {\bibfield  {journal} {\bibinfo
  {journal} {Phys. Rev. A}\ }\textbf {\bibinfo {volume} {86}},\ \bibinfo
  {pages} {015803} (\bibinfo {year} {2012})}\BibitemShut {NoStop}%
\bibitem [{\citenamefont {Grimsmo}\ and\ \citenamefont
  {Parkins}(2013)}]{PhysRevA.87.033814}%
  \BibitemOpen
  \bibfield  {author} {\bibinfo {author} {\bibfnamefont {Arne~L.}\ \bibnamefont
  {Grimsmo}}\ and\ \bibinfo {author} {\bibfnamefont {Scott}\ \bibnamefont
  {Parkins}},\ }\bibfield  {title} {\enquote {\bibinfo {title} {Cavity-qed
  simulation of qubit-oscillator dynamics in the ultrastrong-coupling
  regime},}\ }\href {\doibase 10.1103/PhysRevA.87.033814} {\bibfield  {journal}
  {\bibinfo  {journal} {Phys. Rev. A}\ }\textbf {\bibinfo {volume} {87}},\
  \bibinfo {pages} {033814} (\bibinfo {year} {2013})}\BibitemShut {NoStop}%
\bibitem [{\citenamefont {Zhu}\ \emph {et~al.}(2020)\citenamefont {Zhu},
  \citenamefont {Ping}, \citenamefont {Yang},\ and\ \citenamefont
  {Agarwal}}]{PhysRevLett.124.073602}%
  \BibitemOpen
  \bibfield  {author} {\bibinfo {author} {\bibfnamefont {C.~J.}\ \bibnamefont
  {Zhu}}, \bibinfo {author} {\bibfnamefont {L.~L.}\ \bibnamefont {Ping}},
  \bibinfo {author} {\bibfnamefont {Y.~P.}\ \bibnamefont {Yang}}, \ and\
  \bibinfo {author} {\bibfnamefont {G.~S.}\ \bibnamefont {Agarwal}},\
  }\bibfield  {title} {\enquote {\bibinfo {title} {Squeezed light induced
  symmetry breaking superradiant phase transition},}\ }\href {\doibase
  10.1103/PhysRevLett.124.073602} {\bibfield  {journal} {\bibinfo  {journal}
  {Phys. Rev. Lett.}\ }\textbf {\bibinfo {volume} {124}},\ \bibinfo {pages}
  {073602} (\bibinfo {year} {2020})}\BibitemShut {NoStop}%
\bibitem [{\citenamefont {Schir\'o}\ \emph {et~al.}(2012)\citenamefont
  {Schir\'o}, \citenamefont {Bordyuh}, \citenamefont {\"Oztop},\ and\
  \citenamefont {T\"ureci}}]{PhysRevLett.109.229901}%
  \BibitemOpen
  \bibfield  {author} {\bibinfo {author} {\bibfnamefont {M.}~\bibnamefont
  {Schir\'o}}, \bibinfo {author} {\bibfnamefont {M.}~\bibnamefont {Bordyuh}},
  \bibinfo {author} {\bibfnamefont {B.}~\bibnamefont {\"Oztop}}, \ and\
  \bibinfo {author} {\bibfnamefont {H.~E.}\ \bibnamefont {T\"ureci}},\
  }\bibfield  {title} {\enquote {\bibinfo {title} {Erratum: Phase transition of
  light in cavity qed lattices [phys. rev. lett. 109, 053601 (2012)]},}\ }\href
  {\doibase 10.1103/PhysRevLett.109.229901} {\bibfield  {journal} {\bibinfo
  {journal} {Phys. Rev. Lett.}\ }\textbf {\bibinfo {volume} {109}},\ \bibinfo
  {pages} {229901(E)} (\bibinfo {year} {2012})}\BibitemShut {NoStop}%
\bibitem [{\citenamefont {{Buck}}\ and\ \citenamefont
  {{Sukumar}}(1981)}]{1981PhLA...81..132B}%
  \BibitemOpen
  \bibfield  {author} {\bibinfo {author} {\bibfnamefont {B.}~\bibnamefont
  {{Buck}}}\ and\ \bibinfo {author} {\bibfnamefont {C.~V.}\ \bibnamefont
  {{Sukumar}}},\ }\bibfield  {title} {\enquote {\bibinfo {title} {{Exactly
  soluble model of atom-phonon coupling showing periodic decay and revival}},}\
  }\href {\doibase 10.1016/0375-9601(81)90042-6} {\bibfield  {journal}
  {\bibinfo  {journal} {Physics Letters A}\ }\textbf {\bibinfo {volume} {81}},\
  \bibinfo {pages} {132--135} (\bibinfo {year} {1981})}\BibitemShut {NoStop}%
\bibitem [{\citenamefont {Chen}\ \emph {et~al.}(2020)\citenamefont {Chen},
  \citenamefont {Xie},\ and\ \citenamefont {Chen}}]{PhysRevA.102.063721}%
  \BibitemOpen
  \bibfield  {author} {\bibinfo {author} {\bibfnamefont {Xiang-You}\
  \bibnamefont {Chen}}, \bibinfo {author} {\bibfnamefont {You-Fei}\
  \bibnamefont {Xie}}, \ and\ \bibinfo {author} {\bibfnamefont {Qing-Hu}\
  \bibnamefont {Chen}},\ }\bibfield  {title} {\enquote {\bibinfo {title}
  {Quantum criticality of the rabi-stark model at finite frequency ratios},}\
  }\href {\doibase 10.1103/PhysRevA.102.063721} {\bibfield  {journal} {\bibinfo
   {journal} {Phys. Rev. A}\ }\textbf {\bibinfo {volume} {102}},\ \bibinfo
  {pages} {063721} (\bibinfo {year} {2020})}\BibitemShut {NoStop}%
\bibitem [{\citenamefont {Maciejewski}\ \emph {et~al.}(2015)\citenamefont
  {Maciejewski}, \citenamefont {Przybylska},\ and\ \citenamefont
  {Stachowiak}}]{Maciejewski_2015}%
  \BibitemOpen
  \bibfield  {author} {\bibinfo {author} {\bibfnamefont {Andrzej~J.}\
  \bibnamefont {Maciejewski}}, \bibinfo {author} {\bibfnamefont {Maria}\
  \bibnamefont {Przybylska}}, \ and\ \bibinfo {author} {\bibfnamefont {Tomasz}\
  \bibnamefont {Stachowiak}},\ }\bibfield  {title} {\enquote {\bibinfo {title}
  {An exactly solvable system from quantum optics},}\ }\href {\doibase
  10.1016/j.physleta.2015.03.033} {\bibfield  {journal} {\bibinfo  {journal}
  {Physics Letters A}\ }\textbf {\bibinfo {volume} {379}},\ \bibinfo {pages}
  {1503--1509} (\bibinfo {year} {2015})}\BibitemShut {NoStop}%
\bibitem [{\citenamefont {Eckle}\ and\ \citenamefont
  {Johannesson}(2017)}]{Eckle_2017}%
  \BibitemOpen
  \bibfield  {author} {\bibinfo {author} {\bibfnamefont {Hans-Peter}\
  \bibnamefont {Eckle}}\ and\ \bibinfo {author} {\bibfnamefont {Henrik}\
  \bibnamefont {Johannesson}},\ }\bibfield  {title} {\enquote {\bibinfo {title}
  {A generalization of the quantum rabi model: exact solution and spectral
  structure},}\ }\href {\doibase 10.1088/1751-8121/aa785a} {\bibfield
  {journal} {\bibinfo  {journal} {Journal of Physics A: Mathematical and
  Theoretical}\ }\textbf {\bibinfo {volume} {50}},\ \bibinfo {pages} {294004}
  (\bibinfo {year} {2017})}\BibitemShut {NoStop}%
\bibitem [{\citenamefont {Xie}\ \emph {et~al.}(2017)\citenamefont {Xie},
  \citenamefont {Zhong}, \citenamefont {Batchelor},\ and\ \citenamefont
  {Lee}}]{Xie_2017}%
  \BibitemOpen
  \bibfield  {author} {\bibinfo {author} {\bibfnamefont {Qiongtao}\
  \bibnamefont {Xie}}, \bibinfo {author} {\bibfnamefont {Honghua}\ \bibnamefont
  {Zhong}}, \bibinfo {author} {\bibfnamefont {Murray~T}\ \bibnamefont
  {Batchelor}}, \ and\ \bibinfo {author} {\bibfnamefont {Chaohong}\
  \bibnamefont {Lee}},\ }\bibfield  {title} {\enquote {\bibinfo {title} {The
  quantum rabi model: solution and dynamics},}\ }\href {\doibase
  10.1088/1751-8121/aa5a65} {\bibfield  {journal} {\bibinfo  {journal} {Journal
  of Physics A: Mathematical and Theoretical}\ }\textbf {\bibinfo {volume}
  {50}},\ \bibinfo {pages} {113001} (\bibinfo {year} {2017})}\BibitemShut
  {NoStop}%
\bibitem [{\citenamefont {Xie}\ \emph {et~al.}(2019)\citenamefont {Xie},
  \citenamefont {Duan},\ and\ \citenamefont {Chen}}]{Xie_2019}%
  \BibitemOpen
  \bibfield  {author} {\bibinfo {author} {\bibfnamefont {You-Fei}\ \bibnamefont
  {Xie}}, \bibinfo {author} {\bibfnamefont {Liwei}\ \bibnamefont {Duan}}, \
  and\ \bibinfo {author} {\bibfnamefont {Qing-Hu}\ \bibnamefont {Chen}},\
  }\bibfield  {title} {\enquote {\bibinfo {title} {Quantum rabi-stark model:
  solutions and exotic energy spectra},}\ }\href {\doibase
  10.1088/1751-8121/ab1cf6} {\bibfield  {journal} {\bibinfo  {journal} {Journal
  of Physics A: Mathematical and Theoretical}\ }\textbf {\bibinfo {volume}
  {52}},\ \bibinfo {pages} {245304} (\bibinfo {year} {2019})}\BibitemShut
  {NoStop}%
\bibitem [{\citenamefont {Shen}\ \emph {et~al.}(2022)\citenamefont {Shen},
  \citenamefont {Tang}, \citenamefont {Shi}, \citenamefont {Wu}, \citenamefont
  {Yang},\ and\ \citenamefont {Zheng}}]{PhysRevA.106.023705}%
  \BibitemOpen
  \bibfield  {author} {\bibinfo {author} {\bibfnamefont {Li-Tuo}\ \bibnamefont
  {Shen}}, \bibinfo {author} {\bibfnamefont {Chun-Qi}\ \bibnamefont {Tang}},
  \bibinfo {author} {\bibfnamefont {Zhicheng}\ \bibnamefont {Shi}}, \bibinfo
  {author} {\bibfnamefont {Huaizhi}\ \bibnamefont {Wu}}, \bibinfo {author}
  {\bibfnamefont {Zhen-Biao}\ \bibnamefont {Yang}}, \ and\ \bibinfo {author}
  {\bibfnamefont {Shi-Biao}\ \bibnamefont {Zheng}},\ }\bibfield  {title}
  {\enquote {\bibinfo {title} {Squeezed-light-induced quantum phase transition
  in the jaynes-cummings model},}\ }\href {\doibase
  10.1103/PhysRevA.106.023705} {\bibfield  {journal} {\bibinfo  {journal}
  {Phys. Rev. A}\ }\textbf {\bibinfo {volume} {106}},\ \bibinfo {pages}
  {023705} (\bibinfo {year} {2022})}\BibitemShut {NoStop}%
\bibitem [{\citenamefont {FallasPadilla}\ \emph {et~al.}(2022)\citenamefont
  {FallasPadilla}, \citenamefont {Pu}, \citenamefont {Cheng},\ and\
  \citenamefont {Zhang}}]{PhysRevLett.129.183602}%
  \BibitemOpen
  \bibfield  {author} {\bibinfo {author} {\bibfnamefont {Diego}\ \bibnamefont
  {FallasPadilla}}, \bibinfo {author} {\bibfnamefont {Han}\ \bibnamefont {Pu}},
  \bibinfo {author} {\bibfnamefont {Guo-Jing}\ \bibnamefont {Cheng}}, \ and\
  \bibinfo {author} {\bibfnamefont {Yu-Yu}\ \bibnamefont {Zhang}},\ }\bibfield
  {title} {\enquote {\bibinfo {title} {Understanding the quantum rabi ring
  using analogies to quantum magnetism},}\ }\href {\doibase
  10.1103/PhysRevLett.129.183602} {\bibfield  {journal} {\bibinfo  {journal}
  {Phys. Rev. Lett.}\ }\textbf {\bibinfo {volume} {129}},\ \bibinfo {pages}
  {183602} (\bibinfo {year} {2022})}\BibitemShut {NoStop}%
\bibitem [{\citenamefont {Bakemeier}\ \emph {et~al.}(2012)\citenamefont
  {Bakemeier}, \citenamefont {Alvermann},\ and\ \citenamefont
  {Fehske}}]{PhysRevA.85.043821}%
  \BibitemOpen
  \bibfield  {author} {\bibinfo {author} {\bibfnamefont {L.}~\bibnamefont
  {Bakemeier}}, \bibinfo {author} {\bibfnamefont {A.}~\bibnamefont
  {Alvermann}}, \ and\ \bibinfo {author} {\bibfnamefont {H.}~\bibnamefont
  {Fehske}},\ }\bibfield  {title} {\enquote {\bibinfo {title} {Quantum phase
  transition in the dicke model with critical and noncritical entanglement},}\
  }\href {\doibase 10.1103/PhysRevA.85.043821} {\bibfield  {journal} {\bibinfo
  {journal} {Phys. Rev. A}\ }\textbf {\bibinfo {volume} {85}},\ \bibinfo
  {pages} {043821} (\bibinfo {year} {2012})}\BibitemShut {NoStop}%
\bibitem [{\citenamefont {Hwang}\ \emph {et~al.}(2015)\citenamefont {Hwang},
  \citenamefont {Puebla},\ and\ \citenamefont
  {Plenio}}]{PhysRevLett.115.180404}%
  \BibitemOpen
  \bibfield  {author} {\bibinfo {author} {\bibfnamefont {Myung-Joong}\
  \bibnamefont {Hwang}}, \bibinfo {author} {\bibfnamefont {Ricardo}\
  \bibnamefont {Puebla}}, \ and\ \bibinfo {author} {\bibfnamefont {Martin~B.}\
  \bibnamefont {Plenio}},\ }\bibfield  {title} {\enquote {\bibinfo {title}
  {Quantum phase transition and universal dynamics in the rabi model},}\ }\href
  {\doibase 10.1103/PhysRevLett.115.180404} {\bibfield  {journal} {\bibinfo
  {journal} {Phys. Rev. Lett.}\ }\textbf {\bibinfo {volume} {115}},\ \bibinfo
  {pages} {180404} (\bibinfo {year} {2015})}\BibitemShut {NoStop}%
\bibitem [{\citenamefont {Hwang}\ and\ \citenamefont
  {Plenio}(2016)}]{PhysRevLett.117.123602}%
  \BibitemOpen
  \bibfield  {author} {\bibinfo {author} {\bibfnamefont {Myung-Joong}\
  \bibnamefont {Hwang}}\ and\ \bibinfo {author} {\bibfnamefont {Martin~B.}\
  \bibnamefont {Plenio}},\ }\bibfield  {title} {\enquote {\bibinfo {title}
  {Quantum phase transition in the finite jaynes-cummings lattice systems},}\
  }\href {\doibase 10.1103/PhysRevLett.117.123602} {\bibfield  {journal}
  {\bibinfo  {journal} {Phys. Rev. Lett.}\ }\textbf {\bibinfo {volume} {117}},\
  \bibinfo {pages} {123602} (\bibinfo {year} {2016})}\BibitemShut {NoStop}%
\bibitem [{\citenamefont {Pedernales}\ \emph {et~al.}(2015)\citenamefont
  {Pedernales}, \citenamefont {Lizuain}, \citenamefont {Felicetti},
  \citenamefont {Romero}, \citenamefont {Lamata},\ and\ \citenamefont
  {Solano}}]{Pedernales_2015}%
  \BibitemOpen
  \bibfield  {author} {\bibinfo {author} {\bibfnamefont {J.~S.}\ \bibnamefont
  {Pedernales}}, \bibinfo {author} {\bibfnamefont {I.}~\bibnamefont {Lizuain}},
  \bibinfo {author} {\bibfnamefont {S.}~\bibnamefont {Felicetti}}, \bibinfo
  {author} {\bibfnamefont {G.}~\bibnamefont {Romero}}, \bibinfo {author}
  {\bibfnamefont {L.}~\bibnamefont {Lamata}}, \ and\ \bibinfo {author}
  {\bibfnamefont {E.}~\bibnamefont {Solano}},\ }\bibfield  {title} {\enquote
  {\bibinfo {title} {Quantum rabi model with trapped ions},}\ }\href {\doibase
  10.1038/srep15472} {\bibfield  {journal} {\bibinfo  {journal} {Scientific
  Reports}\ }\textbf {\bibinfo {volume} {5}} (\bibinfo {year} {2015}),\
  10.1038/srep15472}\BibitemShut {NoStop}%
\bibitem [{\citenamefont {Niemczyk}\ \emph {et~al.}(2010)\citenamefont
  {Niemczyk}, \citenamefont {Deppe}, \citenamefont {Huebl}, \citenamefont
  {Menzel}, \citenamefont {Hocke}, \citenamefont {Schwarz}, \citenamefont
  {Garcia-Ripoll}, \citenamefont {Zueco}, \citenamefont {H{\"u}mmer},
  \citenamefont {Solano}, \citenamefont {Marx},\ and\ \citenamefont
  {Gross}}]{Niemczyk2010}%
  \BibitemOpen
  \bibfield  {author} {\bibinfo {author} {\bibfnamefont {T.}~\bibnamefont
  {Niemczyk}}, \bibinfo {author} {\bibfnamefont {F.}~\bibnamefont {Deppe}},
  \bibinfo {author} {\bibfnamefont {H.}~\bibnamefont {Huebl}}, \bibinfo
  {author} {\bibfnamefont {E.~P.}\ \bibnamefont {Menzel}}, \bibinfo {author}
  {\bibfnamefont {F.}~\bibnamefont {Hocke}}, \bibinfo {author} {\bibfnamefont
  {M.~J.}\ \bibnamefont {Schwarz}}, \bibinfo {author} {\bibfnamefont {J.~J.}\
  \bibnamefont {Garcia-Ripoll}}, \bibinfo {author} {\bibfnamefont
  {D.}~\bibnamefont {Zueco}}, \bibinfo {author} {\bibfnamefont
  {T.}~\bibnamefont {H{\"u}mmer}}, \bibinfo {author} {\bibfnamefont
  {E.}~\bibnamefont {Solano}}, \bibinfo {author} {\bibfnamefont
  {A.}~\bibnamefont {Marx}}, \ and\ \bibinfo {author} {\bibfnamefont
  {R.}~\bibnamefont {Gross}},\ }\bibfield  {title} {\enquote {\bibinfo {title}
  {Circuit quantum electrodynamics in the ultrastrong-coupling regime},}\
  }\href {\doibase 10.1038/nphys1730} {\bibfield  {journal} {\bibinfo
  {journal} {Nature Physics}\ }\textbf {\bibinfo {volume} {6}},\ \bibinfo
  {pages} {772--776} (\bibinfo {year} {2010})}\BibitemShut {NoStop}%
\bibitem [{\citenamefont {Forn-D\'{\i}az}\ \emph {et~al.}(2010)\citenamefont
  {Forn-D\'{\i}az}, \citenamefont {Lisenfeld}, \citenamefont {Marcos},
  \citenamefont {Garc\'{\i}a-Ripoll}, \citenamefont {Solano}, \citenamefont
  {Harmans},\ and\ \citenamefont {Mooij}}]{PhysRevLett.105.237001}%
  \BibitemOpen
  \bibfield  {author} {\bibinfo {author} {\bibfnamefont {P.}~\bibnamefont
  {Forn-D\'{\i}az}}, \bibinfo {author} {\bibfnamefont {J.}~\bibnamefont
  {Lisenfeld}}, \bibinfo {author} {\bibfnamefont {D.}~\bibnamefont {Marcos}},
  \bibinfo {author} {\bibfnamefont {J.~J.}\ \bibnamefont {Garc\'{\i}a-Ripoll}},
  \bibinfo {author} {\bibfnamefont {E.}~\bibnamefont {Solano}}, \bibinfo
  {author} {\bibfnamefont {C.~J. P.~M.}\ \bibnamefont {Harmans}}, \ and\
  \bibinfo {author} {\bibfnamefont {J.~E.}\ \bibnamefont {Mooij}},\ }\bibfield
  {title} {\enquote {\bibinfo {title} {Observation of the bloch-siegert shift
  in a qubit-oscillator system in the ultrastrong coupling regime},}\ }\href
  {\doibase 10.1103/PhysRevLett.105.237001} {\bibfield  {journal} {\bibinfo
  {journal} {Phys. Rev. Lett.}\ }\textbf {\bibinfo {volume} {105}},\ \bibinfo
  {pages} {237001} (\bibinfo {year} {2010})}\BibitemShut {NoStop}%
\bibitem [{\citenamefont {Askenazi}\ \emph {et~al.}(2014)\citenamefont
  {Askenazi}, \citenamefont {Vasanelli}, \citenamefont {Delteil}, \citenamefont
  {Todorov}, \citenamefont {Andreani}, \citenamefont {Beaudoin}, \citenamefont
  {Sagnes},\ and\ \citenamefont {Sirtori}}]{Askenazi_2014}%
  \BibitemOpen
  \bibfield  {author} {\bibinfo {author} {\bibfnamefont {B}~\bibnamefont
  {Askenazi}}, \bibinfo {author} {\bibfnamefont {A}~\bibnamefont {Vasanelli}},
  \bibinfo {author} {\bibfnamefont {A}~\bibnamefont {Delteil}}, \bibinfo
  {author} {\bibfnamefont {Y}~\bibnamefont {Todorov}}, \bibinfo {author}
  {\bibfnamefont {L~C}\ \bibnamefont {Andreani}}, \bibinfo {author}
  {\bibfnamefont {G}~\bibnamefont {Beaudoin}}, \bibinfo {author} {\bibfnamefont
  {I}~\bibnamefont {Sagnes}}, \ and\ \bibinfo {author} {\bibfnamefont
  {C}~\bibnamefont {Sirtori}},\ }\bibfield  {title} {\enquote {\bibinfo {title}
  {Ultra-strong light–matter coupling for designer reststrahlen band},}\
  }\href {\doibase 10.1088/1367-2630/16/4/043029} {\bibfield  {journal}
  {\bibinfo  {journal} {New Journal of Physics}\ }\textbf {\bibinfo {volume}
  {16}},\ \bibinfo {pages} {043029} (\bibinfo {year} {2014})}\BibitemShut
  {NoStop}%
\bibitem [{\citenamefont {Chen}\ \emph {et~al.}(2017)\citenamefont {Chen},
  \citenamefont {Wang}, \citenamefont {Li}, \citenamefont {Tian}, \citenamefont
  {Qiu}, \citenamefont {Inomata}, \citenamefont {Yoshihara}, \citenamefont
  {Han}, \citenamefont {Nori}, \citenamefont {Tsai},\ and\ \citenamefont
  {You}}]{PhysRevA.96.012325}%
  \BibitemOpen
  \bibfield  {author} {\bibinfo {author} {\bibfnamefont {Zhen}\ \bibnamefont
  {Chen}}, \bibinfo {author} {\bibfnamefont {Yimin}\ \bibnamefont {Wang}},
  \bibinfo {author} {\bibfnamefont {Tiefu}\ \bibnamefont {Li}}, \bibinfo
  {author} {\bibfnamefont {Lin}\ \bibnamefont {Tian}}, \bibinfo {author}
  {\bibfnamefont {Yueyin}\ \bibnamefont {Qiu}}, \bibinfo {author}
  {\bibfnamefont {Kunihiro}\ \bibnamefont {Inomata}}, \bibinfo {author}
  {\bibfnamefont {Fumiki}\ \bibnamefont {Yoshihara}}, \bibinfo {author}
  {\bibfnamefont {Siyuan}\ \bibnamefont {Han}}, \bibinfo {author}
  {\bibfnamefont {Franco}\ \bibnamefont {Nori}}, \bibinfo {author}
  {\bibfnamefont {J.~S.}\ \bibnamefont {Tsai}}, \ and\ \bibinfo {author}
  {\bibfnamefont {J.~Q.}\ \bibnamefont {You}},\ }\bibfield  {title} {\enquote
  {\bibinfo {title} {Single-photon-driven high-order sideband transitions in an
  ultrastrongly coupled circuit-quantum-electrodynamics system},}\ }\href
  {\doibase 10.1103/PhysRevA.96.012325} {\bibfield  {journal} {\bibinfo
  {journal} {Phys. Rev. A}\ }\textbf {\bibinfo {volume} {96}},\ \bibinfo
  {pages} {012325} (\bibinfo {year} {2017})}\BibitemShut {NoStop}%
\bibitem [{\citenamefont {Forn-D{\'i}az}\ \emph {et~al.}(2017)\citenamefont
  {Forn-D{\'i}az}, \citenamefont {Garc{\'i}a-Ripoll}, \citenamefont
  {Peropadre}, \citenamefont {Orgiazzi}, \citenamefont {Yurtalan},
  \citenamefont {Belyansky}, \citenamefont {Wilson},\ and\ \citenamefont
  {Lupascu}}]{Forn-Diaz2017}%
  \BibitemOpen
  \bibfield  {author} {\bibinfo {author} {\bibfnamefont {P.}~\bibnamefont
  {Forn-D{\'i}az}}, \bibinfo {author} {\bibfnamefont {J.~ ~J.}\ \bibnamefont
  {Garc{\'i}a-Ripoll}}, \bibinfo {author} {\bibfnamefont {B.}~\bibnamefont
  {Peropadre}}, \bibinfo {author} {\bibfnamefont {J.-L.}\ \bibnamefont
  {Orgiazzi}}, \bibinfo {author} {\bibfnamefont {M.~ ~A.}\ \bibnamefont
  {Yurtalan}}, \bibinfo {author} {\bibfnamefont {R.}~\bibnamefont {Belyansky}},
  \bibinfo {author} {\bibfnamefont {C.~ ~M.}\ \bibnamefont {Wilson}}, \ and\
  \bibinfo {author} {\bibfnamefont {A.}~\bibnamefont {Lupascu}},\ }\bibfield
  {title} {\enquote {\bibinfo {title} {Ultrastrong coupling of a single
  artificial atom to an electromagnetic continuum in the nonperturbative
  regime},}\ }\href {\doibase 10.1038/nphys3905} {\bibfield  {journal}
  {\bibinfo  {journal} {Nature Physics}\ }\textbf {\bibinfo {volume} {13}},\
  \bibinfo {pages} {39--43} (\bibinfo {year} {2017})}\BibitemShut {NoStop}%
\bibitem [{\citenamefont {Anappara}\ \emph {et~al.}(2009)\citenamefont
  {Anappara}, \citenamefont {De~Liberato}, \citenamefont {Tredicucci},
  \citenamefont {Ciuti}, \citenamefont {Biasiol}, \citenamefont {Sorba},\ and\
  \citenamefont {Beltram}}]{PhysRevB.79.201303}%
  \BibitemOpen
  \bibfield  {author} {\bibinfo {author} {\bibfnamefont {Aji~A.}\ \bibnamefont
  {Anappara}}, \bibinfo {author} {\bibfnamefont {Simone}\ \bibnamefont
  {De~Liberato}}, \bibinfo {author} {\bibfnamefont {Alessandro}\ \bibnamefont
  {Tredicucci}}, \bibinfo {author} {\bibfnamefont {Cristiano}\ \bibnamefont
  {Ciuti}}, \bibinfo {author} {\bibfnamefont {Giorgio}\ \bibnamefont
  {Biasiol}}, \bibinfo {author} {\bibfnamefont {Lucia}\ \bibnamefont {Sorba}},
  \ and\ \bibinfo {author} {\bibfnamefont {Fabio}\ \bibnamefont {Beltram}},\
  }\bibfield  {title} {\enquote {\bibinfo {title} {Signatures of the
  ultrastrong light-matter coupling regime},}\ }\href {\doibase
  10.1103/PhysRevB.79.201303} {\bibfield  {journal} {\bibinfo  {journal} {Phys.
  Rev. B}\ }\textbf {\bibinfo {volume} {79}},\ \bibinfo {pages} {201303(R)}
  (\bibinfo {year} {2009})}\BibitemShut {NoStop}%
\bibitem [{\citenamefont {Garbe}\ \emph {et~al.}(2017)\citenamefont {Garbe},
  \citenamefont {Egusquiza}, \citenamefont {Solano}, \citenamefont {Ciuti},
  \citenamefont {Coudreau}, \citenamefont {Milman},\ and\ \citenamefont
  {Felicetti}}]{PhysRevA.95.053854}%
  \BibitemOpen
  \bibfield  {author} {\bibinfo {author} {\bibfnamefont {L.}~\bibnamefont
  {Garbe}}, \bibinfo {author} {\bibfnamefont {I.~L.}\ \bibnamefont
  {Egusquiza}}, \bibinfo {author} {\bibfnamefont {E.}~\bibnamefont {Solano}},
  \bibinfo {author} {\bibfnamefont {C.}~\bibnamefont {Ciuti}}, \bibinfo
  {author} {\bibfnamefont {T.}~\bibnamefont {Coudreau}}, \bibinfo {author}
  {\bibfnamefont {P.}~\bibnamefont {Milman}}, \ and\ \bibinfo {author}
  {\bibfnamefont {S.}~\bibnamefont {Felicetti}},\ }\bibfield  {title} {\enquote
  {\bibinfo {title} {Superradiant phase transition in the ultrastrong-coupling
  regime of the two-photon dicke model},}\ }\href {\doibase
  10.1103/PhysRevA.95.053854} {\bibfield  {journal} {\bibinfo  {journal} {Phys.
  Rev. A}\ }\textbf {\bibinfo {volume} {95}},\ \bibinfo {pages} {053854}
  (\bibinfo {year} {2017})}\BibitemShut {NoStop}%
\bibitem [{\citenamefont {Felicetti}\ \emph {et~al.}(2015)\citenamefont
  {Felicetti}, \citenamefont {Pedernales}, \citenamefont {Egusquiza},
  \citenamefont {Romero}, \citenamefont {Lamata}, \citenamefont {Braak},\ and\
  \citenamefont {Solano}}]{PhysRevA.92.033817}%
  \BibitemOpen
  \bibfield  {author} {\bibinfo {author} {\bibfnamefont {S.}~\bibnamefont
  {Felicetti}}, \bibinfo {author} {\bibfnamefont {J.~S.}\ \bibnamefont
  {Pedernales}}, \bibinfo {author} {\bibfnamefont {I.~L.}\ \bibnamefont
  {Egusquiza}}, \bibinfo {author} {\bibfnamefont {G.}~\bibnamefont {Romero}},
  \bibinfo {author} {\bibfnamefont {L.}~\bibnamefont {Lamata}}, \bibinfo
  {author} {\bibfnamefont {D.}~\bibnamefont {Braak}}, \ and\ \bibinfo {author}
  {\bibfnamefont {E.}~\bibnamefont {Solano}},\ }\bibfield  {title} {\enquote
  {\bibinfo {title} {Spectral collapse via two-phonon interactions in trapped
  ions},}\ }\href {\doibase 10.1103/PhysRevA.92.033817} {\bibfield  {journal}
  {\bibinfo  {journal} {Phys. Rev. A}\ }\textbf {\bibinfo {volume} {92}},\
  \bibinfo {pages} {033817} (\bibinfo {year} {2015})}\BibitemShut {NoStop}%
\bibitem [{\citenamefont {Cordeiro}\ \emph {et~al.}(2007)\citenamefont
  {Cordeiro}, \citenamefont {Providência}, \citenamefont {da~Providência},\
  and\ \citenamefont {Nishiyama}}]{Cordeiro_2007}%
  \BibitemOpen
  \bibfield  {author} {\bibinfo {author} {\bibfnamefont {F}~\bibnamefont
  {Cordeiro}}, \bibinfo {author} {\bibfnamefont {C}~\bibnamefont
  {Providência}}, \bibinfo {author} {\bibfnamefont {J}~\bibnamefont
  {da~Providência}}, \ and\ \bibinfo {author} {\bibfnamefont {S}~\bibnamefont
  {Nishiyama}},\ }\bibfield  {title} {\enquote {\bibinfo {title} {The
  buck–sukumar model described in terms of su(2) omites su(1, 1) coherent
  states},}\ }\href {\doibase 10.1088/1751-8113/40/40/010} {\bibfield
  {journal} {\bibinfo  {journal} {Journal of Physics A: Mathematical and
  Theoretical}\ }\textbf {\bibinfo {volume} {40}},\ \bibinfo {pages} {12153}
  (\bibinfo {year} {2007})}\BibitemShut {NoStop}%
\bibitem [{\citenamefont {Cui}\ \emph {et~al.}(2020)\citenamefont {Cui},
  \citenamefont {Gr\'emaud}, \citenamefont {Guo},\ and\ \citenamefont
  {Batrouni}}]{PhysRevA.102.033334}%
  \BibitemOpen
  \bibfield  {author} {\bibinfo {author} {\bibfnamefont {Shifeng}\ \bibnamefont
  {Cui}}, \bibinfo {author} {\bibfnamefont {B.}~\bibnamefont {Gr\'emaud}},
  \bibinfo {author} {\bibfnamefont {Wenan}\ \bibnamefont {Guo}}, \ and\
  \bibinfo {author} {\bibfnamefont {G.~G.}\ \bibnamefont {Batrouni}},\
  }\bibfield  {title} {\enquote {\bibinfo {title} {Nonlinear two-photon
  rabi-hubbard model: Superradiance, photon, and photon-pair bose-einstein
  condensates},}\ }\href {\doibase 10.1103/PhysRevA.102.033334} {\bibfield
  {journal} {\bibinfo  {journal} {Phys. Rev. A}\ }\textbf {\bibinfo {volume}
  {102}},\ \bibinfo {pages} {033334} (\bibinfo {year} {2020})}\BibitemShut
  {NoStop}%
\bibitem [{\citenamefont {Gorlach}\ and\ \citenamefont
  {Poddubny}(2017)}]{PhysRevA.95.053866}%
  \BibitemOpen
  \bibfield  {author} {\bibinfo {author} {\bibfnamefont {Maxim~A.}\
  \bibnamefont {Gorlach}}\ and\ \bibinfo {author} {\bibfnamefont
  {Alexander~N.}\ \bibnamefont {Poddubny}},\ }\bibfield  {title} {\enquote
  {\bibinfo {title} {Topological edge states of bound photon pairs},}\ }\href
  {\doibase 10.1103/PhysRevA.95.053866} {\bibfield  {journal} {\bibinfo
  {journal} {Phys. Rev. A}\ }\textbf {\bibinfo {volume} {95}},\ \bibinfo
  {pages} {053866} (\bibinfo {year} {2017})}\BibitemShut {NoStop}%
\bibitem [{\citenamefont {Mittal}\ \emph {et~al.}(2018)\citenamefont {Mittal},
  \citenamefont {Goldschmidt},\ and\ \citenamefont {Hafezi}}]{Mittal2018}%
  \BibitemOpen
  \bibfield  {author} {\bibinfo {author} {\bibfnamefont {Sunil}\ \bibnamefont
  {Mittal}}, \bibinfo {author} {\bibfnamefont {Elizabeth~A.}\ \bibnamefont
  {Goldschmidt}}, \ and\ \bibinfo {author} {\bibfnamefont {Mohammad}\
  \bibnamefont {Hafezi}},\ }\bibfield  {title} {\enquote {\bibinfo {title} {A
  topological source of quantum light},}\ }\href {\doibase
  10.1038/s41586-018-0478-3} {\bibfield  {journal} {\bibinfo  {journal}
  {Nature}\ }\textbf {\bibinfo {volume} {561}},\ \bibinfo {pages} {502--506}
  (\bibinfo {year} {2018})}\BibitemShut {NoStop}%
\bibitem [{\citenamefont {Olekhno}\ \emph {et~al.}(2020)\citenamefont
  {Olekhno}, \citenamefont {Kretov}, \citenamefont {Stepanenko}, \citenamefont
  {Ivanova}, \citenamefont {Yaroshenko}, \citenamefont {Puhtina}, \citenamefont
  {Filonov}, \citenamefont {Cappello}, \citenamefont {Matekovits},\ and\
  \citenamefont {Gorlach}}]{Olekhno2020}%
  \BibitemOpen
  \bibfield  {author} {\bibinfo {author} {\bibfnamefont {Nikita~A.}\
  \bibnamefont {Olekhno}}, \bibinfo {author} {\bibfnamefont {Egor~I.}\
  \bibnamefont {Kretov}}, \bibinfo {author} {\bibfnamefont {Andrei~A.}\
  \bibnamefont {Stepanenko}}, \bibinfo {author} {\bibfnamefont {Polina~A.}\
  \bibnamefont {Ivanova}}, \bibinfo {author} {\bibfnamefont {Vitaly~V.}\
  \bibnamefont {Yaroshenko}}, \bibinfo {author} {\bibfnamefont {Ekaterina~M.}\
  \bibnamefont {Puhtina}}, \bibinfo {author} {\bibfnamefont {Dmitry~S.}\
  \bibnamefont {Filonov}}, \bibinfo {author} {\bibfnamefont {Barbara}\
  \bibnamefont {Cappello}}, \bibinfo {author} {\bibfnamefont {Ladislau}\
  \bibnamefont {Matekovits}}, \ and\ \bibinfo {author} {\bibfnamefont
  {Maxim~A.}\ \bibnamefont {Gorlach}},\ }\bibfield  {title} {\enquote {\bibinfo
  {title} {Topological edge states of interacting photon pairs emulated in a
  topolectrical circuit},}\ }\href {\doibase 10.1038/s41467-020-14994-7}
  {\bibfield  {journal} {\bibinfo  {journal} {Nature Communications}\ }\textbf
  {\bibinfo {volume} {11}},\ \bibinfo {pages} {1436} (\bibinfo {year}
  {2020})}\BibitemShut {NoStop}%
\bibitem [{\citenamefont {Braak}(2011)}]{PhysRevLett.107.100401}%
  \BibitemOpen
  \bibfield  {author} {\bibinfo {author} {\bibfnamefont {D.}~\bibnamefont
  {Braak}},\ }\bibfield  {title} {\enquote {\bibinfo {title} {Integrability of
  the rabi model},}\ }\href {\doibase 10.1103/PhysRevLett.107.100401}
  {\bibfield  {journal} {\bibinfo  {journal} {Phys. Rev. Lett.}\ }\textbf
  {\bibinfo {volume} {107}},\ \bibinfo {pages} {100401} (\bibinfo {year}
  {2011})}\BibitemShut {NoStop}%
\end{thebibliography}%





\end{document}